\documentclass[10pt,twocolumn,aps,pra,showpacs,nofootinbib,superscriptaddress]{revtex4-1}
\usepackage[english]{babel} 
\usepackage{ifsym}
\usepackage[usenames, dvipsnames]{color} 
\usepackage{graphicx} 
\usepackage{bm}
\usepackage{amsmath} 
\usepackage{amsthm} 
\usepackage{amssymb}
\usepackage{amscd} 
\usepackage{times} 
\usepackage{mathrsfs}
\usepackage[usenames]{color}
\usepackage[pdftex,colorlinks=true,urlcolor=blue, linkcolor=Blue,citecolor=RedViolet]{hyperref}
\newcommand{\hy}[1]{\hyperlink{#1}{\color{Gray} #1}}%my hyperlink
\newcommand{\dd}{{\rm d}}
\newtheorem{thm}{Theorem}
\begin{document}
\title{Mean Value of the Quantum Potential and Uncertainty Relations}
\author{F.~Nicacio}
\email{nicacio@if.ufrj.br} 
\affiliation{Instituto de F\'isica, 
Universidade Federal do Rio de Janeiro, 21941-972, RJ, Brazil.}           
\author{F.~T.~Falciano}
\email{ftovar@cbpf.br} 
\affiliation{CBPF - Brazilian Center for Research in Physics, 
Xavier Sigaud st. 150, zip 22290-180, Rio de Janeiro, RJ, Brazil.}
\affiliation{PPGCosmo, CCE - Federal University of Esp\'\i rito Santo, 
zip 29075-910, Vit\'oria, ES, Brazil.}

%%%%%%%%%%%%%%%%%%%%%%%%%%%%%%%%%%%%%%%%%%%%%%%%%%%%%%%%%%%%%%%%%%%%%%%%%%%%%%%%%%%%%%%%%
\date{\today}
%%%%%%%%%%%%%%%%%%%%%%%%%%%%%%%%%%%%%%%%%%%%%%%%%%%%%%%%%%%%%%%%%%%%%%%%%%%%%%%%%%%%%%%%%
\begin{abstract}
In this work we determine a lower bound to the mean value of the quantum potential for an arbitrary state. Furthermore, we derive a generalized uncertainty relation that is stronger than the Robertson-Schr\"odinger inequality and hence also stronger than the Heisenberg uncertainty principle. The mean value is then associated to the nonclassical part of the covariances of the momenta operator. This imposes a minimum bound for the nonclassical correlations of momenta and gives a physical characterization of the classical and semiclassical limits of quantum systems. The results obtained primarily for pure states are then generalized for density matrices describing mixed states.
\end{abstract}
\maketitle
%%%%%%%%%%%%%%%%%%%%%%%%%%%%%%%%%%%%%%%%%%%%%%%%%%%%%%%%%%%%%%%%%%%%%%%%%%%%%%%%%%%%%%%%%
%%%%%%%%%%%%%%%%%%%%%%%%%%%%%%%%%%%%%%%%%%%%%%%%%%%%%%%%%%%%%%%%%%%%%%%%%%%%%%%%%%%%%%%%%
\section{Introduction}           %%%%%%%%%%%%%%%%%%%%%%%%%%%%%%%%%%%%%%%%%%%%%%%%%%%%%%%%
%%%%%%%%%%%%%%%%%%%%%%%%%%%%%%%%%%%%%%%%%%%%%%%%%%%%%%%%%%%%%%%%%%%%%%%%%%%%%%%%%%%%%%%%%
%%%%%%%%%%%%%%%%%%%%%%%%%%%%%%%%%%%%%%%%%%%%%%%%%%%%%%%%%%%%%%%%%%%%%%%%%%%%%%%%%%%%%%%%%

Quantum mechanics defines a formal procedure to consistently quantize dynamical systems. 
The noncommutability of pairs of operators translates into the well-known uncertainty relations, which is one of the most important kinematic feature of quantum mechanics. However, from a completely different perspective, the debate on interpretation of quantum mechanics frequently focuses on the quantum potential, which seems to have no direct connection with the uncertainty relations due to the lack of an operator definition for it.

In nonrelativistic quantum mechanics, the dynamics is defined by Schr\"odinger equation that unitarily evolves the wave function. Using a polar form for the wave function, Schr\"odinger equation turns into two real coupled equations for the phase and the modulus of the wave function. One of them is very similar to a Hamilton-Jacobi equation for the phase but possessing an extra term, dubbed quantum potential~(\hy{QP}), without a classical analog. The \hy{QP} is responsible for all distinct quantum effects such as entanglement and tunneling. As such, there has been much attention on its properties and several proposals to interpret its physical meaning.

Among the most popular interpretations is Bohmian mechanics, which is a causal interpretation since it dismisses the collapse of the wave function to describe the measurement process~\cite{bohm1952a,bohm1952b,bohm1984,bohm1987,holland1993,Durr1995}. 
The probabilistic description appears due to the unknown initial position of the particle which plays the role of a hidden parameter, hence it is an instantiation of a successful hidden-variable quantum theory in the sense that it reproduces all experimental results of canonical quantum theory. The Born rule, which in this scope is called equilibrium distribution, need not be imposed but can be dynamically derived. It can be shown that initial nonequilibrium states relax to equilibrium on a coarse-grained level\cite{Valentini1990,Valentini2005,Towler2012,Efthymiopoulos2006,Bennett2010}. It is worth mentioning that the ontological nature of the Bohmian trajectories and the interpretation of the \hy{QP} have concrete applications in quantum cosmology \cite{Falciano:2013uaa,Vitenti2012cx,Vitenti2013,PintoNeto2013} and offer a new approach to semi-classical approximations~\cite{Struyve2015,Benseny2014,Struyve2019}.

In Bohmian mechanics the \hy{QP} is interpreted as carrying information but has no material support. Other scenarios give completely different physical interpretation to the \hy{QP}. For instance, in Weyl space~\cite{Carroll2007,Carroll2007b} it is interpreted as a geometrical object associated to the nonmetricity of the metric tensor, hence a manifestation of non-Euclidean geometry at the microscopic scale~\cite{Novello:2011,Falciano2010}, while from the point of view of information theory, a connection with nonrelativistic quantum mechanics appears as a principle of minimum Fisher information~\cite{frieden1995,reginatto1998}. The latter constitutes a rare example of a natural connection between \hy{QP} and uncertainty relations (see~\cite{frieden1988,frieden1989,frieden_1998,reginatto1998,frieden1995,hall2000,hall2001,hall2002} for details).

In the present work we study the mathematical and physical properties of the mean value of the quantum potential (\hy{MVQP}). In contrast to the \hy{QP}, its mean value satisfies inequalities 
that can be used to derive generalized uncertainties relations, which are shown to be more restrictive than the Heisenberg uncertainty principle. Furthermore, the \hy{MVQP} is associated to a parcel of the covariances among all the momenta components, which will be called the nonclassical correlations. Thus, some of our results reproduce part of the Fisher information scenario \cite{frieden1995,reginatto1998} but without including any extra hypothesis. We also depart from this perspective when generalizing the results for mixed states directly from the Liouville von-Neumann equation. 

Our entire analysis is made within the Copenhagen formalism but since it makes no reference to the collapse of the wave function, it can be straightforwardly generalized to other scenarios as well. For instance, the question of the classical and semiclassical limits is described entirely in terms of the presence of nonclassical correlations in the system.

The paper is organized as follows. In the next section we briefly review the basic equations and fix our notation. In sec.~\ref{GURPS} we derive the generalized uncertainty relations for pure states and in sec.~\ref{CCL} we show that the \hy{MVQP} encodes the nonclassical momenta correlations. In sec.~\ref{MxS} we generalize our pure state previous results for density matrices describing mixed states. In sec.~\ref{pmu} we present several comparisons of our results with the Heisenberg and Robertson-Schr\"odinger uncertainties. In sec.~\ref{Exm} we exemplify with concrete physical systems and in sec.~\ref{Concl} we conclude with final remarks.

%%%%%%%%%%%%%%%%%%%%%%%%%%%%%%%%%%%%%%%%%%%%%%%%%%%%%%%%%%%%%%%%%%%%%%%%%%%%%%%%%%%%%%%%%
%%%%%%%%%%%%%%%%%%%%%%%%%%%%%%%%%%%%%%%%%%%%%%%%%%%%%%%%%%%%%%%%%%%%%%%%%%%%%%%%%%%%%%%%%
\section{Classical and Quantum Dynamics} \label{CQD}%%%%%%%%%%%%%%%%%%%%%%%%%%%%%%%%%%%%%%%%%%%%%%%%%
%%%%%%%%%%%%%%%%%%%%%%%%%%%%%%%%%%%%%%%%%%%%%%%%%%%%%%%%%%%%%%%%%%%%%%%%%%%%%%%%%%%%%%%%%
%%%%%%%%%%%%%%%%%%%%%%%%%%%%%%%%%%%%%%%%%%%%%%%%%%%%%%%%%%%%%%%%%%%%%%%%%%%%%%%%%%%%%%%%%

In this section we briefly review some basic equations in order to fix our notation used in the rest of the paper. Let $q:= ( q_1,...,q_n)^\top$ and $ p:=( p_1,..., p_n)^\top$ be column vectors of, 
respectively, the $n$ coordinates and canonically conjugated momenta of a system with $n$ degrees of freedom (\hypertarget{DF}{DF}), which has its evolution governed by 
the Hamiltonian
\begin{equation}\label{classham}
H(q, p) = \tfrac{1}{2} p \cdot {\bf M} p + q \cdot {\bf C} p + p \cdot \xi_p + U(q)\ ,
\end{equation}
where $\xi_p \in \mathbb R^{n}$ is a constant column vector, $ {\bf M} = {\bf M}^\top$ and ${\bf C}$ are $n \times n$ real matrices. The term $U(q)$ is a real function describing any other contribution to the potential energy of the system, such that $H(q, p)$ is the most generic Hamiltonian comprising a quadratic kinetic energy, possibly time-dependent.

Canonical quantum mechanics promote classical variables to operators, hence we have $\hat q:= (\hat q_1,...,\hat q_n)^\dagger$ and $\hat p:=(\hat p_1,...,\hat p_n)^\dagger$ two column vectors of, respectively, coordinates and canonically conjugated momenta operators of the system. Considering the position eigenstates of the system, $| q \rangle := | q_1 \rangle \otimes ... \otimes | q_n \rangle$, the momenta matrix elements are 
\begin{equation}\label{momop}
\langle q'_j |\hat p_k | q_l \rangle = -i \hbar\, \delta(q'_j-q_l) \, \delta_{jl} \, \delta_{lk} \, \partial_k , 
\end{equation}
where $\partial_k := \partial/\partial q_k$. Here we will adopt a symmetric quantization scheme, such that  the quantized version of Hamiltonian (\ref{classham}) becomes the function of operators 
\begin{equation}\label{ham}
H(\hat q,\hat p) = \tfrac{1}{2} \hat p \cdot {\bf M} \hat p + \tfrac{1}{2} \hat q \cdot {\bf C} \hat p + \tfrac{1}{2} \hat p \cdot {\bf C}^\top \hat q + \hat p \cdot \xi_p + U(\hat q)\ .
\end{equation}

In nonrelativistic quantum mechanics the evolution is dictated by Schr\"odinger equation, namely, using the position representation $\psi(q,t) := \langle q |\psi_t\rangle$ we have
\begin{equation}\label{schreq}
i\hbar\partial_t \psi(q,t)=  H\left( q,-i\hbar \partial_q\right) \psi(q,t),
\end{equation}
where $ \partial_q := (\partial_1 ,...,\partial_n )^\top$. As any complex function, the wave function associated to the state $|\psi_t\rangle$ may be written in polar form,  
\begin{equation}\label{wfpolar}
\psi(q,t) = \Omega(q,t) \, {\rm e}^{ \frac{ i }{ \hbar } S(q,t)}\ , 
\end{equation}
where $\Omega(q,t) = |\psi(q,t)|$ and $S(q,t) = \hbar {\rm Arg}[\psi(q,t)]$. Using the polar decomposition for the wave function in the time-dependent Schr\"odinger equation \eqref{schreq}, one obtain two coupled real equations \cite{holland1993} as follows. One is the continuity equation  
\begin{align}\label{classconteq}
\partial_t \Omega^2 + {\partial_q } \cdot \mathcal J_q = 0 
\end{align}
for the probability density $\Omega^2(q,t) = \psi^\ast(q,t)\psi(q,t)$ with probability current given by 
\begin{align} \label{COeq}
\mathcal J_q &:= \Omega^2 \, \partial_p H\big|_{p = \partial_q S} = \Omega^2\!\left( \xi_p + {\bf C}^\top q +{\bf M}\partial_q S \right)\ ;
\end{align}
the other is like the classical Hamilton-Jacobi equation for the phase $S(q,t)$, 
\begin{align}
&\partial_t S + H(q,\partial_q S) + Q(q,t)= 0\ , \label{HJlikeeq}
\end{align}
but with an extra term, 
\begin{equation}
Q(q,t) := -\frac{\hbar^2}{2\Omega} {\partial_q } \cdot {\bf M} {\partial_q \Omega}\ , \label{qp} 
\end{equation}
dubbed the quantum potential (\hypertarget{QP}{QP}), which is a nonlocal potential encoding the information about the state of the system and depends only on $\Omega(q,t)$. Moreover, given its invariance under $\Omega \to k \Omega$ for a constant $k$, we see that the \hy{QP} does not depend on the strength of $\Omega(q,t)$, but only in its form.

In the presence of any sort of classical randomness, the system state in quantum mechanics should be described by a density operator $\hat \rho$, which evolution is governed by the Liouville-von Neumann equation:  $i\hbar \, \partial_t \hat \rho = [\hat H, \hat \rho]$.  Taking the position matrix elements of the evolution equation for the Hamiltonian (\ref{ham}), using a position-completeness relation together with (\ref{momop}), it becomes
\begin{equation}\label{lvnpos}
\!\!i\hbar{\partial_t} \langle q | \hat \rho | {q' } \rangle = \left[ H \left( q, -{i\hbar} \partial_q \right) - H\left( q',  {i\hbar} \partial_{q'} \right)\right] \langle q | \hat \rho | {q' } \rangle.  
\end{equation}

Similarly to (\ref{wfpolar}), we will use the polar decomposition 
\begin{equation}\label{denspol}
\langle q | \hat \rho | {q' } \rangle = \bar\Omega(q, {q' }, t ) \,  \exp\!\left[ \frac{ i }{ \hbar } \bar S(q, {q' }, t )\right]\ ,
\end{equation}
which, when inserted in (\ref{lvnpos}) for the Hamiltonian (\ref{classham}), give us also two coupled differential equations. A continuity like equation that now reads
\begin{equation}\label{COlvn}
\partial_t \bar \Omega^2 + {\partial_q}  \cdot  \mathcal J_{q} +\partial_{q' }  \cdot \mathcal J_{q' }  = 0\ ,   
\end{equation}
where $\mathcal J_{q}$ is exactly written as (\ref{COeq}) but replacing $\Omega(q,t) \to \bar \Omega(q,q',t)$ and $S(q,t) \to \bar S(q,q',t)$ and the current associated to the $q'$ coordinates is 
\[
\mathcal J_{q'} := \bar\Omega^2\!\left( \xi_p + {\bf C}^\top {q' }  -{\bf M}\partial_{q'} \bar S  \right)\ .  
\]

The other equation is again a kind of Hamilton-Jacobi equation with some extra terms:
\begin{align} \label{HJlvn}
\partial_t \bar S &+ H(q,\partial_q \bar S) -\frac{\hbar^2}{2\bar\Omega} {\partial_q} \cdot \mathbf{M} \partial_q \bar\Omega \,   \nonumber \\                                                                            
&- H(q',-\partial_{q'} \bar S) + \frac{\hbar^2}{2\bar\Omega} \partial_{q' } \cdot \mathbf{M} \partial_{q'} \bar\Omega = 0\ .
\end{align}

For a pure state $\hat \rho = |\psi_t\rangle \! \langle \psi_t|$, $\bar \Omega (q,q',t) = \Omega(q,t) \Omega(q',t)$, and $\bar S(q,q',t) = S(q,t) - S(q',t)$. In this circumstance, one can apply a separation of variables into the partial differential equations (\ref{COlvn}) and (\ref{HJlvn}) to obtain, respectively, two versions of (\ref{classconteq}) and (\ref{HJlikeeq}); the separation constant, possibly a function of $t$, 
can be regarded as a shift of the potential energy.

%%%%%%%%%%%%%%%%%%%%%%%%%%%%%%%%%%%%%%%%%%%%%%%%%%%%%%%%%%%%%%%%%%%%%%%%%%%%%%%%%%%%%%%%%
%%%%%%%%%%%%%%%%%%%%%%%%%%%%%%%%%%%%%%%%%%%%%%%%%%%%%%%%%%%%%%%%%%%%%%%%%%%%%%%%%%%%%%%%%

%%%%%%%%%%%%%%%%%%%%%%%%%%%%%%%%%%%%%%%%%%%%%%%%%%%%%%%%%%%%%%%%%%%%%%%%%%%%%%%%%%%%%%%%%
%%%%%%%%%%%%%%%%%%%%%%%%%%%%%%%%%%%%%%%%%%%%%%%%%%%%%%%%%%%%%%%%%%%%%%%%%%%%%%%%%%%%%%%%%
\section{Quantum Potential Uncertainty Relations for Pure States} \label{GURPS}  %%%%%%%%
%%%%%%%%%%%%%%%%%%%%%%%%%%%%%%%%%%%%%%%%%%%%%%%%%%%%%%%%%%%%%%%%%%%%%%%%%%%%%%%%%%%%%%%%%
%%%%%%%%%%%%%%%%%%%%%%%%%%%%%%%%%%%%%%%%%%%%%%%%%%%%%%%%%%%%%%%%%%%%%%%%%%%%%%%%%%%%%%%%%
We consider the amplitude $\Omega: {\mathbb R}^n \times {\mathbb R} \to {\mathbb R}_+$ 
of the pure state $|\psi\rangle$ in (\ref{wfpolar}) as a (classical) probability density 
function twice differentiable and continuous everywhere in $\mathbb R^{n+1}$. 
Since, by definition, it is nonnegative and 
\[
\int_{\mathbb R^n} \!\!\!\!\! d^n \! q \,\, 
[\Omega(q,t)]^2  = 1,  \,\,\, \forall t \in\mathbb R\ ,
\]
thus $\lim_{||q|| \to \infty} \Omega(q,t) = 0$; this excludes nonnormalizable solutions 
of the Schr\"odinger equation to provide good candidates for $\Omega$. 

The set of all square-integrable functions with respect to the measure $\Omega^2$ is denoted as $L^2(\Omega^2)$. We also assume that any element of $L^2(\Omega^2)$ is continuous and has continuous first and second derivatives. If $T_i, T_j \in L^2(\Omega^2)$, then the mean-value and the covariances of these functions are defined, respectively, by
\begin{equation} \label{cov}
\begin{split} 
& \langle T_i \rangle := \int_{\mathbb R^n} \!\!\!\!\! d^n \! q \,\,  [\Omega(q,t)]^2 \, T_i(q)\ , \\ 
&{\rm Cov} (T_i,T_j) := \langle T_i T_j \rangle - \langle T_i \rangle \langle T_j \rangle\ . 
\end{split}
\end{equation}
Furthermore, by the Cauchy-Schwartz inequality \cite{kulkarni1999} we have
\begin{equation} \label{csi}  
|{\rm Cov} (T_i,T_j)|^2 \le {\rm Cov}(T_i,T_i) \, {\rm Cov}(T_j,T_j)\ .
\end{equation}
As a matter of compactness, we shall write for a vector function $T: {\mathbb R}^n \to {\mathbb R}^n$ and ${\mathbf {\text Cov}}(T,T)$ means the $n \times n$ matrix with elements ${\rm Cov}({T}_i,{T}_j)$ for $i,j = 1,...n$.

The mean value of the quantum potential (\hypertarget{MVQP}{MVQP}) can readily be obtained from Eq.(\ref{qp}) and reads
\begin{equation} \label{meanqp}
\begin{split} 
\langle Q(t) \rangle &= -\frac{\hbar^2}{2} \int_{{\mathbb R}^n} \!\!\!\!\!\! d^n\! q \,\, \Omega \,\, \partial_q \cdot {\bf M} {\partial_q \Omega}          \\
&= \frac{\hbar^2}{8} \int_{{\mathbb R}^n} \!\!\!\!\!\! d^n\! q \,\, \partial_q {\rm ln}\Omega^2 \cdot {\bf M} \partial_q \Omega^2 \ , 
\end{split}
\end{equation}
where we omit the $(q,t)$-variables in $\Omega$, which is responsible for the time dependence of $\langle Q(t) \rangle$. Furthermore, it is constrained by the following theorem.

\begin{thm}\label{thmQ}
Let a quantum system with $n$ degrees of freedom have its evolution governed by the Hamiltonian \eqref{ham} where the kinetic matrix ${\bf M}$ is {positive definite}, real, and symmetric. 
If the system is in a pure state, given a generic function $T_0 \in L^2(\Omega^2)$, the mean value of the quantum potential given by \eqref{meanqp} satisfies the following inequality
\begin{equation}\label{qpur}
\langle Q(t) \rangle \ge {\mathcal L}_Q(T_0,t):= \frac{\hbar^2}{8} 
\frac{ \left\langle {\partial_q T_0}\right\rangle \cdot{\bf M} \left\langle {\partial_q T_0}\right\rangle}{ {\rm Cov}(T_0,T_0) } \ . 
\end{equation}
\end{thm}

\textit{Proof.}--- Since the matrix ${\bf M}$ is real and symmetric, it can be diagonalized by a real orthogonal transformation: ${\bf M}  = {\bf O^{\!\top}\!\! \Lambda O}$, where ${\bf \Lambda} := {\rm Diag} (\lambda_1,...,\lambda_n)$ is the diagonal matrix of the positive real eigenvalues of ${\bf M}$, {\it i.e.} $\lambda_i > 0\, \forall i$. Thus, we can define the functions 
\begin{equation}  \label{auxfunc}
T_i (q) := \sum_{j,k =1}^{n} \delta_{ij}  \sqrt{\lambda_j} {\bf O}_{jk} \, \partial_k {\rm ln}\Omega^2 \,\,\, (i = 1,...,n)\ , 
\end{equation}
all of which belonging to $L^2(\Omega^2)$. Note that $\langle T_i \rangle = 0$ for $i = 1,...,n$, since $\Omega^2 \to 0$ as $||q|| \to \infty$. Rewriting the \hy{MVQP} in (\ref{meanqp}) using (\ref{auxfunc}), one finds
\begin{equation}  \label{meanqp3}
\langle Q(t) \rangle = \frac{\hbar^2}{8} \sum_{i=1}^{n} {\rm Cov}(T_i,T_i)\ .
\end{equation}

Using the definition \eqref{cov}, the covariance of $T_0 \in L^2(\Omega^2)$  with each $T_i$ in (\ref{auxfunc}) is given by
\[
{\rm Cov}(T_0,T_i) = - \sum_{j,k =1}^{n} \delta_{ij} \sqrt{\lambda_j} {\bf O}_{jk} \left\langle {\partial_k T_0}\right\rangle\ .
\]
Therefore, squaring and summing for all functions $T_i$ in \eqref{auxfunc}, 
\begin{equation}  \label{covrel}
\sum_{i = 1}^n |{\rm Cov}(T_0,T_i)|^2 = \left\langle{\partial_q T_0} \right\rangle \cdot{\bf M}  \left\langle{\partial_q T_0} \right\rangle .
\end{equation}
Finally, using the Cauchy-Schwartz inequality \eqref{csi} together with \eqref{meanqp3}-\eqref{covrel} we obtain \eqref{qpur}.  \hfill$\square$\\

Given that ${{\rm Cov}(T_0,T_0)} \ge 0$ and ${\bf M}$ is a positive-definite matrix, the arbitrariness of the function $T_0$ in (\ref{qpur}) implies that $\langle Q(t)\rangle > 0$, which is a remarkable and, as far as we know, new property of the \hy{QP}. Furthermore, the {\it bound function} in (\ref{qpur}) is affine symmetric, namely ${\mathcal L}_Q(\alpha T_0 + \beta,t) = {\mathcal L}_Q(T_0, t)$ for $\alpha, \beta \in \mathbb R$.

In quantum mechanics, the uncertainty relations \cite{Sakurai1994,Cohen1977} are related with pairs of noncommuting quantum operators. In particular, the Robertson-Schr\"odinger uncertainty relation \cite{Robertson1929,Schrodinger1930}, derived from the canonical position-momentum commutation relation, plays a central role. Note however that the \hy{QP} inequality is completely different. Theorem~\ref{thmQ} shows that the \hy{QP} satisfies the inequality \eqref{qpur} for any function $T_0 \in L^2(\Omega^2)$. In principle, one can choose all kind of functions to relate to the \hy{QP}: 
this raises a multitude of possible inequalities in (\ref{qpur}). Despite the derivation relies on classical probability rules, this kind of generalized uncertainty relation is associated to the (quantum) randomness of the system. We shall analyze these characteristics in detail in the following sections but now we want to prove another important result. 

\begin{thm}\label{thmbQ}
Let a quantum system with $n$ degrees of freedom have its evolution governed by the Hamiltonian \eqref{ham} where the kinetic matrix ${\bf M}$ is  {positive definite}, real and symmetric. If the system is in a pure state, there is a specific function $T_\ast \in L^2(\Omega^2)$ that extremizes the bound on the mean value of the quantum potential given by \eqref{qpur} such that the inequality depends only on $\Omega^2$ and will be given by
\begin{equation} \label{qpur2}
\langle Q(t) \rangle \ge \frac{\hbar^2}{8} \, \lambda_\wedge \!\left({\bf Q} \right)\ ,
\end{equation}
where $\lambda_\wedge$ is the largest eigenvalue of the $n \times n$ real matrix ${\bf Q}$ defined as 
\begin{equation} \label{meanqp2}
{\mathbf Q} := - \left \langle {\partial^2_{qq} {\rm ln} \Omega^2} \right\rangle {\bf M}\ .
\end{equation}
\end{thm}

\textit{Proof.}--- ${\mathcal L}_Q(T_0, t)$ can be viewed as a functional of $T_0(q)$ and let us suppose that it has at least one extremum at $T_0(q) = T_\ast(q)$. Consider small variations around this function as
\begin{equation}   \label{var1}
\widetilde T_0(q) = T_\ast(q) + \epsilon\, \phi(q) \,\,\, \text{with} \,\,\, \epsilon \ll 1\ , 
\end{equation}
where $\phi(q) \in L^2(\Omega^2)$ is a continuous and differentiable function. Keeping only first order terms in $\epsilon$ and imposing $\delta {\mathcal L}_Q := {\mathcal L}_Q(\widetilde T_0,t)- {\mathcal L}_Q(T_\ast,t) = 0$ we find
\[
\frac{ {\rm Cov}(T_\ast,T_\ast) } { \left\langle{\partial_q T_\ast} \right\rangle	\cdot{\bf M}  	\left\langle{\partial_q T_\ast}\right\rangle } =
\frac{ {\rm Cov}(T_\ast,\phi) }{ \left\langle{\partial_q T_\ast} \right\rangle	\cdot{\bf M}  \left\langle{\partial_q \phi}\right\rangle }\ , 
\]%end{equation}
which can be recast as
\[
\begin{split}
&\!\!\!\!\!\! \int_{\mathbb R^n} \!\!\!\!\! d^n \! q \,\, \Omega^2 \phi
\left[ {\partial_q {\rm ln}\Omega^2}\, \cdot{\bf M}  \left\langle {\partial_q T_\ast}\right\rangle {\rm Cov}(T_\ast,T_\ast)  \right] +                           \\
&\!\!\!\!\!\! \int_{\mathbb R^n} \!\!\!\!\! d^n \! q \,\, \Omega^2 \phi \left[ \left\langle{\partial_q T_\ast} \right\rangle  \cdot{\bf M}  \left\langle {\partial_q T_\ast} \right\rangle (T_\ast - \langle T_\ast\rangle)\right] = 0 \, .
\end{split}
\]
Given the arbitrariness of $\phi(q)$ in (\ref{var1}) we conclude that 
\begin{equation}    \label{extbound}
T_\ast(q) =  \langle T_\ast\rangle - \frac{ \hbar^2 }{ 8 } \frac{\partial_q  {\rm ln}\Omega^2 \cdot{\bf M} \left\langle \partial_q T_\ast \right\rangle}{{\mathcal L}_Q(T_\ast,t)}\ .
\end{equation}

Taking the derivative of (\ref{extbound}) with respect to $q$ and averaging with $\Omega^2$ over all state space, we have
\begin{equation} \label{extbound2}
\frac{\hbar^2}{8} {\bf Q} \left\langle {\partial_q T_\ast} \right\rangle= {\mathcal L}_Q  \left\langle {\partial_q T_\ast} \right\rangle\ ,
\end{equation}
where ${\bf Q}$ is the $n \times n$ real matrix defined in \eqref{meanqp2}. Thus, the extreme of ${\mathcal L}_Q(T_0,t)$ is an eigenvalue of ${\bf Q}$ associated to the eigenvector $\left\langle {\partial_q T_\ast} \right\rangle$. Therefore, the functional in (\ref{qpur}) is bounded by 
\[
\hbar^2\lambda_\vee ({\bf Q})/8 \le {\mathcal L}_Q(T_0,t) \le \hbar^2\lambda_\wedge ({\bf Q})/8\ ,
\]
where $\lambda_\wedge$ (resp. $\lambda_\vee$) is the largest (resp. smallest) eigenvalue of the matrix ${\bf Q}$. Since $\langle Q(t) \rangle >0$, the largest eigenvalue in $\lambda_\wedge \!\left({\bf Q} \right)$ must be positive. In order to obtain the most constrained bound possible, we can choose the largest eigenvalue, namely $\lambda_\wedge \!\left({\bf Q} \right)$.  \hfill$\square$\\

Given a quantum state with probability amplitude $\Omega(q,t)$, the solution $T_\ast$ of equation (\ref{extbound}) is a necessary and sufficient condition for the extremization of ${\mathcal L}_Q(T_0,t)$~\cite{gelfand1963}. In other words, assuming that it exists, the extremum of ${\mathcal L}_Q(T_0,t)$ must satisfy Eq.(\ref{extbound}). This extremum will be a maximum for specific functions $\Omega(q,t)$, which will be now explored.

%%%%%%%%%%%%%%%%%%%%%%%%%%%%%%%%%%%%%%%%%%%%%%%%%%%%%%%%%%%%%%%%%%%%%%%%%%%%%%%%%%%%%%%%%
%%%%%%%%%%%%%%%%%%%%%%%%%%%%%%%%%%%%%%%%%%%%%%%%%%%%%%%%%%%%%%%%%%%%%%%%%%%%%%%%%%%%%%%%%
\subsection*{Sufficient Conditions for the Maximum: 
Gaussian States and the Linear Function}        \label{scfe} %%%%%%%%%%%%%%%%%%%%%%%%%%%%
%%%%%%%%%%%%%%%%%%%%%%%%%%%%%%%%%%%%%%%%%%%%%%%%%%%%%%%%%%%%%%%%%%%%%%%%%%%%%%%%%%%%%%%%%
%%%%%%%%%%%%%%%%%%%%%%%%%%%%%%%%%%%%%%%%%%%%%%%%%%%%%%%%%%%%%%%%%%%%%%%%%%%%%%%%%%%%%%%%%
We will analyze particular solutions of (\ref{extbound}) such as to 
construct sufficient conditions for extrema of the function $\mathcal L_{Q}$.

\ %%%%%%%%%%%%%%%%%%%%%%%%%%%%%%%%%%%%%%%%%%%%%%%%%%%%%%%%%%%%%%%%%%%%%%%%%%%%%%%%%%%%%%%%%
\paragraph{Linear Bound Function:}\label{lbf} %%%%%%%%%%%%%%%%%%%%%%%%%%%%%%%%%%%%%%%%%%%%%
%%%%%%%%%%%%%%%%%%%%%%%%%%%%%%%%%%%%%%%%%%%%%%%%%%%%%%%%%%%%%%%%%%%%%%%%%%%%%%%%%%%%%%%%%
The simplest nontrivial inequality in (\ref{qpur}) is attained for the linear 
function $T_0(q) = \zeta \cdot q + \zeta_0 \in L^2(\Omega^2)$, 
with $\zeta \in \mathbb R^n$ and $\zeta_0 \in \mathbb R$ two constant vectors. 
Inserting this in (\ref{qpur}), 
\[
{\mathcal L}_{Q}(\zeta\cdot q + \zeta_0, t) = 
\frac{\hbar^2}{8} \frac{ \zeta \cdot{\bf M} \zeta }{ \zeta \cdot {\mathbf V}   \zeta }\ ,
\]
where ${\bf V} = {\textbf{Cov}}(q,q) > 0$
is the the position covariance matrix (\hy{PCM}) defined through (\ref{cov})%
\footnote{The matrix 
${\mathbf {\text Cov}}(q,q) = \left\langle qq^{\!\top} \right\rangle - 
\left\langle q \right\rangle \! \left\langle q \right\rangle^{\!\top} > 0$ 
is the same as the one obtained for a pure state $\hat \rho = |\psi\rangle \! \langle \psi|$, 
when inserting position completeness relations in (\ref{cm}) and using the polar structure (\ref{wfpolar}). 
Note also that a null eigenvalue of $\bf V$ would imply total precision of a position measurement, 
which is forbidden by the Heisenberg uncertainty principle, 
thus the positive definiteness of ${\textbf{Cov}}(q,q)$.}.  
Note that the above ${\mathcal L}_Q$ is a relative Rayleigh quotient, hence, by the Courant-Fischer theorem \cite{horn2013}, 
\begin{equation} \label{linbound}
\frac{\hbar^2}{8} \lambda_\vee( {\mathbf V}^{-1}{\bf M} ) \le {\mathcal L}_{Q}(\zeta\cdot q + \zeta_0, t)  \le \frac{\hbar^2}{8} \lambda_\wedge({\mathbf V}^{-1}{\bf M})\ ,
\end{equation}
where $\lambda_\vee$ (resp. $\lambda_\wedge$) is the smallest (resp. largest) eigenvalue of ${\bf M} {\mathbf V}^{-1}$ and the equality occurs when $\zeta = \zeta_\vee$ (resp. $\zeta = \zeta_\wedge$) is the eigenvector associated to $\lambda_\vee$ (resp. $\lambda_\wedge$).

Since (\ref{qpur}) is valid for any function $T_0(q) \in L^2(\Omega^2)$, we can write 
\begin{eqnarray} \label{qpurlf}
\langle Q(t) \rangle \ge {\mathcal L}_{Q}(\zeta\cdot q + \zeta_0, t) := \frac{\hbar^2}{8} \, \lambda_\wedge \left({\bf V}^{-1}{\bf M} \right)\ .
\end{eqnarray}

As long as ${\bf V}$ and ${\bf M}$ are positive definite symmetric matrices, the above eigenvalue is positive. The limiting interval in (\ref{linbound}) and the bound in (\ref{qpurlf}) for the function ${\mathcal L}_{Q}$ are valid for any quantum state and depend on it only through its covariance matrix $\bf V$. Notwithstanding, nothing inhibits that another choice of $T_0(q)$ will provide a greater (better) bound for the \hy{MVQP}. Thus, the linear function constitutes only a sufficient condition for (\ref{linbound}) and (\ref{qpurlf}).

\ %%%%%%%%%%%%%%%%%%%%%%%%%%%%%%%%%%%%%%%%%%%%%%%%%%%%%%%%%%%%%%%%%%%%%%%%%%%%%%%%%%%%%%%
\paragraph{Gaussian States:}%%%%%%%%%%%%%%%%%%%%%%%%%%%%%%%%%%%%%%%%%%%%%%%%%%%%%%%%%%%%%
%%%%%%%%%%%%%%%%%%%%%%%%%%%%%%%%%%%%%%%%%%%%%%%%%%%%%%%%%%%%%%%%%%%%%%%%%%%%%%%%%%%%%%%%%
The probability amplitude for a generic pure Gaussian state is given by 
[see \eqref{wfgauss}-\eqref{ampph}]
\begin{equation} \label{gauspol}
\Omega(q,t) = 
\frac{ \exp\left[-\frac{1}{4} (q - \eta_q) \cdot {\bf V}^{-1}(q - \eta_q)\right]  }
{\left[  (2\pi)^n \det {\bf V} \right]^{1/4}}\ ,
\end{equation}
where ${\bf V}$ is the position covariance matrix and 
$\eta_q := \langle \psi | \hat q | \psi \rangle$ is the position vector of the mean values. 
For the amplitude of a Gaussian state in (\ref{gauspol}), 
the solution of (\ref{extbound}) is a linear function, {\it i.e.}, 
$T_\ast(q) = \zeta \cdot q + \zeta_0$, where, according to \eqref{extbound2}, 
$\zeta$ is one of the eigenvectors of the matrix ${\bf Q} = {\bf V}^{-1}{\bf M}$ [see \eqref{meanqp2}]. If we choose the largest eigenvalue, then $T_0(q) = \zeta_\wedge \cdot q + \zeta_0$ is a necessary condition to 
\[
{\mathcal L}_Q(T_0,t) \le {\mathcal L}_Q(\zeta_\wedge \cdot q + \zeta_0,t) = \frac{\hbar^2}{8}\lambda_\wedge ( {\bf V}^{-1}{\bf M})\ . 
\]

We have previously shown that the linear function is a sufficient condition for \eqref{linbound}, which is valid for arbitrary states. Thus, a linear function $T_0(q) = \zeta_\wedge \cdot q + \zeta_0$, where $\zeta_\wedge$ is the eigenvector associated to the largest eigenvalue of the matrix ${\bf V}^{-1}{\bf M}$, is a necessary and sufficient condition for a maximum value of the functional $\mathcal{L}_{Q}(T_0(q),t)$ when the state of the system is Gaussian\footnote{It is important to take into account that any requirement of the state to be Gaussian can be relaxed to a state with a Gaussian probability density (\ref{gauspol}), since the phase of such states does not play any role in our results, {\it i.e.}, the phase of the state does not necessarily have the quadratic form described in (\ref{ampph}).}. 

In conclusion, the linear function is the solution that extremizes 
the bound function $\mathcal{L}_{Q}(T_0(q),t)$ 
when the state is Gaussian and {\it vice-versa}, 
namely, the set of states that has the linear function as the solution 
that extremizes $\mathcal{L}_{Q}(T_0(q),t)$ consists of Gaussian states.
%
%%%%%%%%%%%%%%%%%%%%%%%%%%%%%%%%%%%%%%%%%%%%%%%%%%%%%%%%%%%%%%%%%%%%%%%%%%%%%%%%%%%%%%%%%
%%%%%%%%%%%%%%%%%%%%%%%%%%%%%%%%%%%%%%%%%%%%%%%%%%%%%%%%%%%%%%%%%%%%%%%%%%%%%%%%%%%%%%%%%
\section{Correlations and The Classical Limit} \label{CCL} %%%%%%%%%%%%%%%%%%%%%%%%%%%%%%
%%%%%%%%%%%%%%%%%%%%%%%%%%%%%%%%%%%%%%%%%%%%%%%%%%%%%%%%%%%%%%%%%%%%%%%%%%%%%%%%%%%%%%%%%
%%%%%%%%%%%%%%%%%%%%%%%%%%%%%%%%%%%%%%%%%%%%%%%%%%%%%%%%%%%%%%%%%%%%%%%%%%%%%%%%%%%%%%%%%
For a generic mixed or pure state  $\hat \rho$, we define the $n \times n$ 
(symmetric) {\it position covariance matrix} (\hypertarget{PCM}{PCM}) and 
the $n \times n$ (symmetric) {\it momentum covariance matrix} (\hypertarget{MCM}{MCM}), respectively, as 
\begin{equation} 
\begin{split}   \label{cm}              
{\mathbf V} &:= \text{Tr} \left( \hat \rho \, \hat q \hat q^\dag \right)  - \text{Tr} \left( \hat \rho \, \hat q \right) \text{Tr} \left( \hat \rho \, \hat q^\dag \right)\ , \\
\widetilde{\mathbf V} &:= \text{Tr} \left( \hat \rho \, \hat p \hat p^\dag \right)  -  \text{Tr} \left( \hat \rho \, \hat p \right) \text{Tr} \left( \hat \rho \, \hat p^\dag \right)\ . 
\end{split} 
\end{equation}
In this section we will continue to deal only with pure states, $\hat \rho = |\psi\rangle\!\langle \psi|$, 
and postpone the appropriate generalization for mixed states to the subsequent section.

Using the wave function of a pure state in the polar form, Eq.(\ref{wfpolar}), the correlations contained in the \hy{MCM} of the state can be broken into two distinct contributions. In fact, inserting position-com\-ple\-te\-ness relations into (\ref{cm}), considering the matrix elements in (\ref{momop}), and using (\ref{denspol}), it is possible to show that
\begin{equation} \label{momcorr}
\widetilde{\mathbf V} = \widetilde{\mathbf V}_{\!\text{c}} +  \widetilde{\mathbf V}_{\!\text{nc}}\ , 
\end{equation}
where 
\begin{equation} \label{correl}
\widetilde{\mathbf V}_{\!\text{c}}  := {\mathbf{Cov}}\!\left({\partial_q S},{\partial_q S}\right), \,\,\, 
\widetilde{\mathbf V}_{\!\text{nc}}  :=  \left\langle - \frac{\hbar^2}{\Omega}\partial^2_{qq} \Omega \right\rangle\ , 
\end{equation}
with the mean value and the covariance both defined in (\ref{cov}), {\it i.e.}, using the ``classical'' probability $\Omega^2$.

The above structure distinguishes the correlations $\widetilde{\mathbf V}_{\!\text{c}}$ generated by the Hamilton-Jacobi dynamics \eqref{HJlikeeq} and the purely quantum ones in $\widetilde{\mathbf V}_{\!\text{nc}}$. This description is in accordance with the notion developed in~\cite{hall2001,hall2002}, where the momenta operator is decomposed as a sum of a classical and a nonclassical operators, $\hat p = \hat p_{\text{c}}+\hat p_{\text{nc}}$.  The decomposition is such that the mean value of the classical part is
\begin{equation} \label{clmom}
p_{\text{c}} := \langle \psi | \hat p | \psi \rangle  = \left \langle \partial_q S \right\rangle\ . 
\end{equation}
As a consequence, the nonclassical operator, while having null mean value,  $p_{\text{nc}} := \langle \psi | \hat p_\text{nc} | \psi \rangle = 0$, still influences the system dynamics due to ``quantum induced noises'' through the nonclassical correlations represented by $\widetilde{\mathbf V}_{\text{nc}}$. Note that this matrix is related to the concavity of the function $\Omega(q,t)$, actually to a kind of ``mean concavity''. 

Comparing the first line in (\ref{meanqp}) with $\widetilde{\mathbf V}_{\!\text{nc}}$ in Eq.(\ref{correl}), one finds 
\begin{equation}  \label{meanqpcorr}
\langle Q(t) \rangle = \frac{1}{2} {\rm Tr} \left[ \widetilde{\mathbf V}_{\!\text{nc}} {\bf M}\right]\ ,
\end{equation}
which shows that \hy{MVQP} can be interpreted as a measure of the ``quantumness'' of the state of the system inasmuch as the Hamilton-Jacobi equation gives the classical dynamics. We can obtain the same result by integrating by parts \eqref{meanqp2}, thus the matrix $\bf Q$ [defined in \eqref{meanqp2}] can be written as 
\begin{equation}  \label{nonclascorr}
\frac{\hbar^2}{4} {\bf Q } = \widetilde{\mathbf V}_{\!\text{nc}}{\bf M}\ .
\end{equation}

Therefore, the inequality for the \hy{MVQP} on (\ref{qpur2}) sets a bound on the quantum and classical correlations, which are constrained by the uncertainty relation
\begin{equation}\label{nccorur}
{\rm Tr} \left[ \widetilde{\mathbf V}_{\!\text{nc}} {\bf M}\right] = {\rm Tr} \left[ \widetilde{\mathbf V}{\bf M} -  \widetilde{\mathbf V}_{\!\text{c}} {\bf M} \right] \ge \lambda_\wedge \!\left(\widetilde{\mathbf V}_{\!\text{nc}}{\bf M} \right)\ .
\end{equation}

Since ${\bf M} > 0$, the inertias%
%%%%%%%%%%%%%%%%%%%%%%%%%%%%%%%%%%%%%%%%%%%%%%%%
\footnote{The {\it inertia} is the triple containing the number of positive, the number of negative, and the number of null eigenvalues of a matrix \cite{horn2013}.} % 
%%%%%%%%%%%%%%%%%%%%%%%%%%%%%%%%%%%%%%%%%%%%%%%%
of $\bf Q$ and $\widetilde{\mathbf V}_{\!\text{nc}}$ are the same \cite{horn2013}, even though in principle generic. Notwithstanding, the relation \eqref{nccorur} imposes a stronger physical constraint: whatever the sign of the eigenvalues of $\widetilde{\mathbf V}_{\!\text{nc}}$, the quantum potential uncertainty relation guarantees a minimum of quantum correlations determined by the largest (positive) eigenvalue of $\bf Q$ or $\widetilde{\mathbf V}_{\!\text{nc}}$.

A quantum system approaches the classical limit {\it where} the \hy{QP} is negligible. However, the \hy{QP} can vanish only in some regions of the configuration space, since we have proven that 
$\langle Q(t) \rangle > 0$. The latter is a statement about the average over the whole configuration space and the positivity condition of the \hy{MVQP} by itself is not sufficient to forbid the classical behavior of the system. In fact, the vanishing of $\langle Q(t) \rangle$ would imply the vanishing of the nonclassical correlations $\widetilde{\mathbf V}_{\!\text{nc}}$ due to the positive definiteness of ${\bf M}$, see \eqref{meanqpcorr}. Therefore, \eqref{nccorur} can be understood as saying that it is impossible to find a quantum state that has no quantum momenta correlations.

In addition, the semiclassical limit is commonly taken as the rough limit $\hbar \rightarrow 0$. A WKB approximation consists in keeping only first order terms in $\hbar$, which, in principle, succeeds to describe all sorts of quantum phenomena such as superposition, entanglement and coherence. Thus, it is not clear what is discarded when we neglect second or higher order terms in $\hbar$. In contrast, using our description, the situation is more precise. In terms of the correlations, the WKB approximation describes quantum systems that are dominated by classical correlations, {\it i.e.}, the nonclassical ones are small compared to the classical correlations. For instance, the \hy{QP} of a Gaussian state such as \eqref{meangaussqp} does not vanish in the semiclassical limit, since ${\bf V}$ depends on $\hbar$, see (\ref{complexcov2}). This is consistent with the fact that not all pure Gaussian states are WKB wave packets \cite{maia2008}.

As a last comment, from the structure of Eqs.(\ref{momcorr}) and (\ref{correl}), the decomposition of the momenta is such that $\langle \psi | \hat p_{\text{nc}} \, \hat p_{\text{c}}^\dagger | \psi \rangle = 0$, 
{\it i.e.}, the classical and nonclassical components are linearly uncorrelated, namely, they do describe independent degrees of freedom. 

%%%%%%%%%%%%%%%%%%%%%%%%%%%%%%%%%%%%%%%%%%%%%%%%%%%%%%%%%%%%%%%%%%%%%%%%%%%%%%%%%%%%%%%%%
%%%%%%%%%%%%%%%%%%%%%%%%%%%%%%%%%%%%%%%%%%%%%%%%%%%%%%%%%%%%%%%%%%%%%%%%%%%%%%%%%%%%%%%%%
\section{Mixed States} \label{MxS}%%%%%%%%%%%%%%%%%%%%%%%%%%%%%%%%%%%%%%%%%%%%%%%%%%%%%%%
%%%%%%%%%%%%%%%%%%%%%%%%%%%%%%%%%%%%%%%%%%%%%%%%%%%%%%%%%%%%%%%%%%%%%%%%%%%%%%%%%%%%%%%%%
%%%%%%%%%%%%%%%%%%%%%%%%%%%%%%%%%%%%%%%%%%%%%%%%%%%%%%%%%%%%%%%%%%%%%%%%%%%%%%%%%%%%%%%%%
In the last section we analyzed only pure states. 
Now we proceed to generalize all previous results to mixed states evolving under the 
Liouville-von Neumann equation (\ref{lvnpos}). In equation (\ref{HJlvn}), 
derived from (\ref{lvnpos}), there are two analogous terms 
to the quantum potential defined in (\ref{qp}). 
One of them is
\begin{equation} \label{mixedqp}
\bar Q(q,q',t) := -\frac{\hbar^2}{2\bar\Omega} {\partial_q} \cdot \mathbf{M} \partial_q \bar\Omega\ 
\end{equation}
and the other is equal to $-\bar Q(q',q,t)$. All the results in this section are invariant under such interchanges between $q'$ and $q$, since $\bar\Omega(q,q',t) = \bar\Omega(q',q,t)$. Hence, it will be enough to work with definition  (\ref{mixedqp}). As will become clear soon, it will be enough for our purposes to deal with the quantity $\bar Q(q,q',t)\big|_{q'=q}$, which means that we calculate the function in (\ref{mixedqp}) and only afterwards evaluate the diagonal terms by making $q'=q$. 

As before, we interpret $\bar\Omega(q,q'=q,t)$ in (\ref{denspol}) as a (classical) probability, since $\bar\Omega: {\mathbb R}^{2n+1} \to {\mathbb R}$ is non-negative and 
\[
\int_{\mathbb R^{n}} \!\!\!\!\! \dd^n \! q \,\, \bar \Omega(q,q,t)  = 1\ ,  \,\,\, \forall t \in\mathbb R\ .
\]
Again we assume that $\bar \Omega(q,q',t)$ is twice differentiable and continuous everywhere in $\mathbb R^{2n+1}$. Consequently, 
\[
\lim_{||q|| \to \infty} \bar \Omega(q,q',t) = \lim_{||q'|| \to \infty} \bar \Omega(q,q',t) = 0\ .
\]
Note, however, that $\bar \Omega(q,q',t)$ is a probability density only when $q'=q$. In order to make a clear distinction from (\ref{cov}), the (ensemble) mean value of a function $\bar T_i \in L^2(\bar \Omega)$ will be denote with a sub-index $\rho$ as 
\[
\langle \bar T_i \rangle_{\rho} := \int_{\mathbb R^n} \!\!\!\!\! d^n \! q \,\,  \Omega(q,q,t) \, \bar T_i(q)\ ,
\]
and similarly for the covariance, which will be denoted by ${\rm Cov} (\bar T_i,\bar T_j)_{\rho}$.

Calculating the mean value of (\ref{mixedqp}) with respect to the probability measure $\bar \Omega(q,q,t)$ we get
\begin{equation}                                                                     \label{meanqpmix}
\begin{split}
\langle \bar Q (t)\rangle_{\rho} &=  \int_{\mathbb R^n} \!\!\!\!\! \dd^n \! q \,\,  \bar\Omega(q,q,t) \, \bar Q(q,q',t)\big|_{q'=q}                    \\                    
&=  -\frac{\hbar^2}{2}\!\int_{\mathbb R^n} \!\!\!\!\! \dd^n \! q \,\,  \partial_q \cdot {\bf M} \partial_q \bar \Omega\big|_{q'=q}\ .
\end{split}
\end{equation}

Inserting the completeness relation in position on the definition (\ref{cm}) and using the polar structure (\ref{denspol}), it is not difficult to show that $\widetilde{\mathbf V}$ still decomposes as $\widetilde{\mathbf V}=\widetilde{\mathbf V}_{\! \text{c}}+\widetilde{\mathbf V}_{\! \text{nc}}$ but now with 
\begin{equation}                                                                         \label{correlmix}
\begin{split}
\widetilde{\mathbf V}_{\! \text{c}} & :=   {\mathbf{Cov}}\! \left({\partial_q \bar S\big|_{q'=q}},{\partial_q \bar S\big|_{q'=q}}\right)_{\!\rho},            \\
\widetilde{\mathbf V}_{\! \text{nc}} & :=   \left\langle \left[ - \frac{\hbar^2}{\bar\Omega}\partial^2_{qq} \bar\Omega\right]_{q'=q} \right\rangle_{\!\rho}\ .
\end{split}
\end{equation}
The above definitions for mixed states are natural extensions of the ``quantum-classical'' dichotomy of the momenta operator, where the mean value of the nonclassical part $\hat p_\text{nc}$ is null, {\it i.e.}, $p_\text{nc} = 0$, while the classical part is such that 
\begin{equation}\label{clmomixed}
p_{\text c} := \text{Tr} \left( \hat \rho \, \hat p \right) = \left \langle \partial_{q} \bar S \big|_{q'=q} \right\rangle_{\!\rho}\ .
\end{equation}

As expected, both equations in (\ref{correlmix}) and (\ref{clmomixed}) reduce, respectively, to (\ref{correl}) and (\ref{clmom}) for pure states $\hat \rho = |\psi \rangle \! \langle \psi |$. In addition, one has $\bar \Omega(q,q',t) = \Omega(q,t)\Omega(q',t)$ and $\bar S(q,q',t) = S(q,t) - S(q',t)$.

An interesting result is that, for a generic mixed state, with the above definitions, the relation between the \hy{MVQP} and the nonclassical correlations of momenta is still preserved. Indeed, it is easy to see from Eqs.(\ref{meanqpmix}) and (\ref{correlmix}) that
\begin{equation}                                                                         \label{mixqp}
\langle \bar Q (t)\rangle_{\rho} = \frac{1}{2} {\rm Tr} \left[ \widetilde{\mathbf V}_{\!\text{nc}} {\bf M}\right]\ .
\end{equation}

Since $\hat \rho$ is a positive definite operator operator with unity trace, we can choose its spectral decomposition
\begin{equation}                                                                         \label{spectral}
\hat \rho = \sum_k \omega_k | \psi_k \rangle \! \langle \psi_k|,  \,\,\, \sum_k \omega_k = 1\ ,
\end{equation}
where $\omega_k \ge 0$ and $| \psi_k \rangle$ are, respectively, its eigenvalues and eigenvectors. Expression \eqref{spectral} is called a convex decomposition of $\hat \rho$ in terms of pure states. Inserting (\ref{spectral}) in (\ref{cm}), one recovers the well known result that a convex combination of covariance matrices is also a covariance matrix (see, for instance, \cite{Cox1974}). In fact, (\ref{clmomixed}) with the decomposition (\ref{spectral}) reads  
\[
p_{\text c} := \text{Tr} \left( \hat \rho \, \hat p \right) =  \sum_k \omega_k \left \langle \partial_{q} S_k \right\rangle, \,\,\, \left \langle \partial_{q} S_k \right\rangle = \langle \psi_k | \hat p | \psi_k \rangle\ .
\]
From (\ref{cm}), which also can be written as $\widetilde{\mathbf V} = \sum_k \omega_k \langle \psi_k | (\hat p-p_{\text c})(\hat p-p_{\text c})^\dag | \psi_k \rangle$, we find 
\[
\widetilde{\mathbf V} = \sum_k \omega_k \left[ \widetilde{\mathbf V}_{\! \text{c}}^{(k)}  +  \widetilde{\mathbf V}_{\! \text{nc}}^{(k)} +  \delta{\mathbf V}^{(k)} \right]\ ,
\]
where 
\begin{equation} \label{correlmix2}
\!\widetilde{\mathbf V}_{\! \text{c}}^{(k)} := {\mathbf{Cov}}\!\left({\partial_q S_k},{\partial_q S_k}\right), \,\, \widetilde{\mathbf V}_{\! \text{nc}}^{(k)} :=  \left\langle\! - \frac{\hbar^2}{\Omega_k}\partial^2_{qq} \Omega_k \!\right\rangle
\end{equation}
are the matrices in (\ref{correl}) for each eigenstate $|\psi_k\rangle$ of the decomposition and 
\begin{equation}\label{correlmix3}
{\delta \bf V}^{(k)} :=  \left\langle \partial_q S_k - p_{\text c} \right\rangle \left\langle \partial_q S_k - p_{\text c} \right\rangle^{\!\top}
\end{equation}
are the correlations induced by the statistical mixture. To obtain (\ref{correlmix2}) and (\ref{correlmix3}), we wrote each state of the decomposition as (\ref{wfpolar}), with  $\Omega_k(q,t) = |\langle q|\psi_k\rangle|$ and 
$S_k(q,t) = \hbar \, {\rm arg} (\langle q|\psi_k\rangle)$. 

Neither the phase $\bar S(q,q',t)$ nor the amplitude $\bar \Omega(q,q',t)$ is a convex combination, respectively, of the phases and amplitudes of the pure states $|\psi_k\rangle$. Thus, due to the terms (\ref{correlmix3}), the matrices $\widetilde{\mathbf V}_{\! \text{c}}$ and $\widetilde{\mathbf V}_{\! \text{nc}}$ in (\ref{correlmix}) are not decomposable exclusively into convex combinations of $\widetilde{\mathbf V}_{\! \text{c}}^{(k)}$ and $\widetilde{\mathbf V}_{\! \text{nc}}^{(k)}$.

However, $\widetilde{\mathbf V}_{\! \text{nc}}$ in (\ref{correlmix}) still can be written as a convex sum, just rewriting properly the second derivatives of $\bar \Omega(q,q',t)$. This {\it tour de force} is carefully detailed in Appendix \ref{cdV} and the final form is
\begin{equation}                                                                         \label{qcconv}
\widetilde{\mathbf V}_{\! \text{nc}} = \sum_{k}\omega_k \widetilde{\mathbf V}_{\! \text{nc}}^{(k)} +  
\delta \widetilde{\bf V}_{\! \text{nc}}\ ,
\end{equation}
where $\delta \widetilde{\bf V}_{\! \text{nc}}$ is the symmetric positive-semidefinite matrix given by (\ref{apB7}). 
Now, we are in a position to establish a generalized uncertainty relation, analogous to \eqref{qpur2}, for mixed states.

\begin{thm}\label{thmbQmix}
Let a quantum system with $n$ degrees of freedom to have its evolution governed by the Hamiltonian \eqref{ham} where the kinetic matrix ${\bf M}$ is  {positive definite}, real, and symmetric. 
If the system is in a mixed state, the \hy{MVQP} defined in \eqref{meanqpmix} has a lower bound given by
\begin{equation}\label{qpmix2}
\langle \bar Q (t)\rangle_{\rho} \ge \frac{\hbar^2}{8} \sum_{k} \omega_k \, \lambda_\wedge \!\left({{\bf V}^{(k)}}^{-1} {\bf M} \right)\ ,
\end{equation}
where $\lambda_\wedge$ is the largest eigenvalue of ${{\bf V}^{(k)}}^{-1} {\bf M}$ and ${\bf V}^{(k)}$ is the \hy{PCM} defined in (\ref{cm}) for each state $|\psi_k\rangle$.
\end{thm}

\textit{Proof.}--- Using \eqref{mixqp} and \eqref{qcconv}, the \hy{MVQP} reads
\begin{equation}                                                                         \label{qpmix}
\begin{split}
\langle \bar Q (t)\rangle_{\rho} &= \frac{1}{2} \sum_{k} \omega_k {\rm Tr} \left[ \widetilde{\mathbf V}_{\!\text{nc}}^{(k)} {\bf M} \right] + \frac{1}{2} {\rm Tr} \left[ \delta \widetilde{\mathbf V}_{\!\text{nc}} {\bf M} \right] \\
& \ge \frac{1}{2} \sum_{k} \omega_k {\rm Tr} \left[ \widetilde{\mathbf V}_{\!\text{nc}}^{(k)} {\bf M} \right] = \sum_{k} \omega_k \left\langle Q_k (t)\right\rangle\ , 
\end{split}
\end{equation}
where we have used the fact that  $\bf M$ is positive-definite and $\delta \widetilde{\mathbf V}_{\!\text{nc}}$ is positive semidefinite. Each $\left\langle Q_k (t)\right\rangle$ is the \hy{MVQP} for each pure state of the convex decomposition (\ref{spectral}), and each one of them is bounded by a respective function $\mathcal L_{Q}^k(T_0^k,t)$ in (\ref{qpur}). By choosing linear functions $T_0^k(q) = \zeta_{\wedge}^k \cdot q + \zeta_0^k$, where $\zeta_{\wedge}^k \in \mathbb{R}^n$ is the eigenvector associated to the largest eigenvalue $\lambda_\wedge$ of ${{\bf V}^{(k)}}^{-1} {\bf M}$, and $\zeta_0^k \in \mathbb{R}$ is a constant, we immediately arrive at \eqref{qpmix2}. \hfill$\square$\\

Similarly to the pure case, a quantum system in a mixed state has a minimum of quantum correlations. As long as $\delta \widetilde{\bf V}_{\! \text{nc}}$ is a symmetric positive-semidefinite matrix, the decomposition \eqref{qcconv} shows that 
\begin{equation}                                                                         \label{nccorurmix}
\begin{split}
{\rm Tr} \left[ \widetilde{\mathbf V}_{\!\text{nc}} {\bf M} \right] &\ge \sum_{k} \omega_k \, \lambda_\wedge \!\left(\widetilde{\mathbf V}_{\!\text{nc}}^{(k)}{\bf M} \right) \\ 
&\ge \lambda_\wedge \!\left( \sum_{k} \omega_k \widetilde{\mathbf V}_{\!\text{nc}}^{(k)}{\bf M} \right)\ ,
\end{split}
\end{equation}
where the last inequality relies on Weil's theorem for the sum of eigenvalues \cite{horn2013}. 

%%%%%%%%%%%%%%%%%%%%%%%%%%%%%%%%%%%%%%%%%%%%%%%%%%%%%%%%%%%%%%%%%%%%%%%%%%%%%%%%%%%%%%%%%
%%%%%%%%%%%%%%%%%%%%%%%%%%%%%%%%%%%%%%%%%%%%%%%%%%%%%%%%%%%%%%%%%%%%%%%%%%%%%%%%%%%%%%%%%
\section{Position-Momentum Uncertainties}  \label{pmu}                       %%%%%%%%%%%%
%%%%%%%%%%%%%%%%%%%%%%%%%%%%%%%%%%%%%%%%%%%%%%%%%%%%%%%%%%%%%%%%%%%%%%%%%%%%%%%%%%%%%%%%%
%%%%%%%%%%%%%%%%%%%%%%%%%%%%%%%%%%%%%%%%%%%%%%%%%%%%%%%%%%%%%%%%%%%%%%%%%%%%%%%%%%%%%%%%%
In this section we present several comparisons of our results with the Heisenberg and Robertson-Schr\"odinger uncertainty principles.  To this end, we need the position-momentum covariance matrix,  which is the $2n \times 2n$ symmetric and positive-definite matrix defined through the following block structure 
\[
\mathcal V := \left( \begin{array}{ll}
                     {\bf V} & {\bf V}_{\! qp} \\
                     {\bf V}_{\! qp}^\top & \widetilde{\bf V} 
                     \end{array} \right)\ , 
\]
where ${\bf V}$ and $\widetilde{\bf V}$ are the $n\times n$ symmetric and positive-definite matrices in (\ref{cm}). The $n\times n$ matrix ${\bf V}_{\! qp}$ encodes 
the covariances among positions and momenta: 
\[
{\bf V}_{\! qp} :=  
\text{Tr}\left(  \hat \rho \, 
               \{\!\!\{ \hat q,\hat p  \}\!\!\} \right) - 
               \{\!\!\{ \text{Tr} ( \hat \rho \,\hat q ) , \text{Tr} ( \hat \rho \, \hat p ) \}\!\!\} \ , 
\]
where $\{\!\!\{ A, B\}\!\!\} := \tfrac{1}{2}( A B^\dag +  B A^\dag)$. The Robertson-Schr\"odinger uncertainty relation (\hypertarget{RSUR}{RSUR}) is written as \cite{PhysRevLett.84.2726}
\[
\mathcal V + \frac{i \hbar}{2}\mathsf J \ge 0, 
\]
where $\mathsf J$ is defined in (\ref{sympdyn}). Since ${\bf V} > 0$, the above condition on $\mathcal V$ can be expressed in terms of the Schur complement \cite{horn2013}
\begin{equation}                                                                         \label{schur}
\widetilde{\bf V} - 
\left[ {\bf V}_{\! qp} + \tfrac{i\hbar}{2}\mathsf I_n \right]^\dag 
{\bf V}^{-1}  
\left[ {\bf V}_{\! qp} + \tfrac{i\hbar}{2}\mathsf I_n \right] \ge 0\ . 
\end{equation}

For a system with only one-\hy{DF},   
\begin{equation}                                                                         \label{rs1d}
\Delta q^2 \Delta p^2 - [{\rm Cov}(q,p)]^2 - \frac{\hbar^2}{4} \ge 0\ ,  
\end{equation}
which is a sufficient condition to the Heisenberg principle $\Delta q \Delta p \ge \frac{\hbar}{2}$. Let us now compare our results with the \hy{RSUR} for separate cases: an arbitrary one-\hy{DF} system; pure Gaussian states; quantum states with no classical correlations; and a system with $n$ independent \hy{DF}.

\ %%%%%%%%%%%%%%%%%%%%%%%%%%%%%%%%%%%%%%%%%%%%%%%%%%%%%%%%%%%%%%%%%%%%%%%%%%%%%%%%%%%%%%% 
\paragraph{Systems with one DF:} %%%%%%%%%%%%%%%%%%%%%%%%%%%%%% 
%%%%%%%%%%%%%%%%%%%%%%%%%%%%%%%%%%%%%%%%%%%%%%%%%%%%%%%%%%%%%%%%%%%%%%%%%%%%%%%%%%%%%%%%% 
Considering a system with only one-\hy{DF} and described by a pure state, we write ${\bf M} = 1/m > 0$ in (\ref{ham}) and the general uncertainty relation (\ref{qpur}) becomes 
\begin{eqnarray}                                                                         \label{qpur1d}
\langle Q(t) \rangle \ge {\mathcal L}_Q(T_0,t):= \frac{\hbar^2}{8m} 
\frac{ \left\langle{\partial_q T_0}\right\rangle^2 }
     { {\rm Cov}(T_0,T_0) }\ .
\end{eqnarray}

The maximum value attained by the function $\mathcal L_Q$ in \eqref{extbound2} simplifies to 
\begin{equation}                                                                         \label{extbound21d}
{\mathcal L}_Q(T_\ast(q),t) = - \frac{\hbar^2}{8m} \left \langle 
                 \partial^2_{qq} {\rm ln} \Omega^2
                 \right\rangle\ .
\end{equation}

From the definition of the \hy{MVQP} in (\ref{meanqp2}), 
it is evident that $\left\langle Q(t) \right\rangle = {\mathcal L}_Q(T_\ast(q),t)$ for one dimensional systems. Actually, the equality is a direct consequence of the saturation of the Cauchy-Schwartz inequality in (\ref{csi}). This inequality becomes an equality if and only if the functions $T_i(q)$ and $T_j(q)$ are linearly correlated \cite{kulkarni1999}, {\it i.e.}, when there exists $\alpha,\beta,\gamma \in \mathbb R$, such that $\alpha T_i(q) + \beta T_j(q) + \gamma = 0$. Accordingly, the saturation of the uncertainty relation for the \hy{MVQP} in (\ref{qpur}) occurs when such a linear relation is obeyed by $T_i$ in (\ref{auxfunc}) and $T_0$. 

When a system described by a pure state has only one-\hy{DF}, there exists only one $T_i(q)$ in (\ref{auxfunc}) and it is easy to see from (\ref{extbound}) that 
\[
T_\ast(q) = \langle T_\ast \rangle - 
\frac{\hbar^2}{8} 
\frac{\langle \partial_q T_\ast\rangle}{\mathcal L_Q(T_\ast,t)} \, T_1(q)\ .
\]

This means that, for all one-\hy{DF} systems, the uncertainty principle in (\ref{qpur})  is saturated by the solution $ T_0(q) = T_\ast(q)$. Therefore, in general, the saturation of the \hy{MVQP} for mixed states differs from the sum of the \hy{MVQP} for each pure state comprising the mixed state (see (\ref{qpmix})). As a last comment, as far as we know, there is not a relation between the saturation of the \hy{RSUR} and the dimension of the system. 

\ %%%%%%%%%%%%%%%%%%%%%%%%%%%%%%%%%%%%%%%%%%%%%%%%%%%%%%%%%%%%%%%%%%%%%%%%%%%%%%%%%%%%%%% 
%\paragraph{The QP Uncertainty for One DF:}%%%%%%%%%%%%%%%%%%%%%%%%%%%%%%%%%%%%%%%%%%%%%%% 
%%%%%%%%%%%%%%%%%%%%%%%%%%%%%%%%%%%%%%%%%%%%%%%%%%%%%%%%%%%%%%%%%%%%%%%%%%%%%%%%%%%%%%%%% 

Even though there is no new information about the behavior of the quantum potential in \eqref{extbound21d}, the inequality in (\ref{qpur1d}) is still valid and can give important information about the system. In particular, the generalized uncertainty relation is stronger than the \hy{RSUR} as shown by the following theorem.

\begin{thm}\label{thmsRSUR}
Let a quantum system with one-degree of freedom to have its evolution governed by the Hamiltonian \eqref{ham} where the kinetic matrix ${\bf M}$ is {positive definite}, real, and symmetric.  If the system is in a pure state, its variance on position $\Delta q^2$ and momentum $\Delta p^2$ satisfies the inequality
\begin{equation}                                                                         \label{lur1df2}
\Delta p^2 \Delta q^2 
- {\rm Cov}\left({\partial_q S},{\partial_q S}\right)\Delta q^2
-\frac{\hbar^2}{4} \ge 0\ , 
\end{equation}
where $S(q,t)$ is the phase of the wave function describing the state of the system. Furthermore, inequality \eqref{lur1df2} is stronger than the \hy{RSUR} in the sense that is a sufficient but not necessary condition for
the \hy{RSUR}.
\end{thm}

\textit{Proof.}--- For the system considered, we can choose a linear function 
$T_0(q) = \zeta q + \zeta_0$, with $\zeta,\zeta_0 \in \mathbb R$, 
to obtain the one-\hy{DF} version of (\ref{qpurlf}):
\begin{equation}                                                                         \label{lur1df}
\langle Q(t) \rangle \, \Delta q^2 \ge \frac{\hbar^2 }{8m}\ .
\end{equation}

In addition, in this case the covariance matrix is simply the position variance, ${\mathbf V} = \Delta q^2$ while the \hy{MVQP} is only related to the nonclassical correlations, see (\ref{meanqpcorr}). The one-\hy{DF} version of (\ref{momcorr}), where $\widetilde{\mathbf V} = \Delta p^2$, $\widetilde{\mathbf V}_{\!\text{c}} = {\rm Cov}\left({\partial_q S},{\partial_q S}\right)$, and $\widetilde{\mathbf V}_{\!\text{nc}} = \Delta p^2_{\text{nc}}$, when inserted in (\ref{lur1df}), gives exactly \eqref{lur1df2}. In order to prove that it is stronger than the \hy{RSUR}, let us define the quantity ${\bf \delta} := {\rm Cov}\left({\partial_q S},{\partial_q S}\right)\Delta q^2 -[{\rm Cov}(q,p)]^2$. Summing it to both sides of (\ref{lur1df2}), we recognize its LHS identical to the LHS of Eq.(\ref{rs1d}), while the RHS is just ${\bf \delta}$. Defining the vector $x := (\partial_q S,q)^\top$, one notes that ${\bf \delta} = \det {\rm \bf Cov}(x,x)$. Since a generic covariance matrix is always non-negative \cite{kulkarni1999}, then ${\bf \delta} \ge 0$. \hfill$\square$\\

The above result shows that the generalized uncertainty principle for the quantum potential (\ref{lur1df}) is a sufficient condition to the \hy{RSUR} (\ref{schur}), for any one-\hy{DF} system.

\ %%%%%%%%%%%%%%%%%%%%%%%%%%%%%%%%%%%%%%%%%%%%%%%%%%%%%%%%%%%%%%%%%%%%%%%%%%%%%%%%%%%%%%% 
\paragraph{Pure Gaussian States:}    %%%%%%%%%%%%%%%%%%%%%%%%%%%%%%%%%%%%%%%%%%%%%%%%%% 
%%%%%%%%%%%%%%%%%%%%%%%%%%%%%%%%%%%%%%%%%%%%%%%%%%%%%%%%%%%%%%%%%%%%%%%%%%%%%%%%%%%%%%%
For a generic pure Gaussian state, the \hy{QP} and its mean value, using (\ref{meanqp}), read
\begin{eqnarray} 
Q(q,t) &=& \tfrac{\hbar^2}{4} \,  {\rm Tr} \! \left( {\bf V}^{-1}{\bf M} \right) \nonumber \\ 
&-& \tfrac{\hbar^2}{8} (q - \eta_q) \cdot {\bf V}^{-1}{\bf M} {\bf V}^{-1} (q - \eta_q)\ ,\nonumber\\
\langle Q(t) \rangle &=& \tfrac{\hbar^2}{8} \, {\rm Tr} \! \left( {\bf V}^{-1}{\bf M} \right)\ . \label{meangaussqp}       
\end{eqnarray}
It is interesting to notice that the \hy{MVQP} 
is half of the maximum value of the \hy{QP}, which is reached for the center of the wavepacket 
$\langle q \rangle = \eta_q$, {\it i.e.}, $\langle Q(t)\rangle = \tfrac{1}{2} Q(\eta_q,t)$.

We have shown in Sec.\ref{scfe} that the system in a pure Gaussian state is a sufficient condition for the maximization of the bound function in (\ref{qpur}). Thus, we can find a relation between the correlations of the system that are more restrictive than the \hy{RSUR}. Using the definition (\ref{meanqp2})  for the matrix $\bf Q$ together with the amplitude for a pure Gaussian state (\ref{gauspol}), the identification (\ref{nonclascorr}) can be recast as
\begin{equation}                                                                         \label{urpart}
{\bf V} \widetilde{\mathbf V}_{\!\text{nc}} = \frac{\hbar^2}{4} \mathsf I_{n}\ ,
\end{equation}
which establishes an exact relation (instead of an inequality) between the quantum and classical correlations for the states. For the one-\hy{DF} case, it reduces to   
\begin{equation}                                                                         \label{qp1dfur}
\Delta p_{\text nc}^2 \Delta q^2 = \frac{\hbar^2}{4}\ ,
\end{equation}
which can be seen as an equality encoded within Heisenberg's uncertainty relation. Note that \eqref{qp1dfur} is simply (\ref{lur1df}) for the present case.

Noteworthy, the uncertainty relation in (\ref{urpart}) is stronger than the \hy{RSUR} for pure Gaussian states. Actually, we will prove that (\ref{urpart}) is a sufficient condition to (\ref{schur}). We begin by using (\ref{momcorr}) in (\ref{urpart}), and choosing $w \in \mathbb C^n$ in order to write an inner product as 
\[ 
w^\dag\left[ \widetilde{\mathbf V} - \widetilde{\mathbf V}_{\!\text{c}} - \frac{\hbar^2}{4} {\bf V}^{-1} \right] w = 0 \ .
\]

Adding the term ${\bf \Delta} :=\widetilde{\mathbf V}_{\!\text{c}} - 
{\bf V}_{\! qp}^\top{\bf V}^{-1}{\bf V}_{\! qp}$ on both sides of the above equation, and noting that the pair of Hermitian conjugated terms $(\frac{i\hbar}{2}{\bf V}_{\! qp}^\top {\bf V}^{-1}, -  \frac{i\hbar}{2}{\bf V}^{-1}{\bf V}_{\! qp})$ give zero contribution to the inner product, we find
\[
 w^\dag \left[ \widetilde{\bf V} - \left( {\bf V}_{\! qp} + \tfrac{i\hbar}{2}\mathsf I_n \right)^{\!\dag} {\bf V}^{-1}  
\left({\bf V}_{\! qp} + \tfrac{i\hbar}{2}\mathsf I_n \right) \right] w   \ge w^\dag {\bf \Delta} w\ ,
\]
where the matrix inside the brackets is the same as in (\ref{schur}). 

Note that $\bf \Delta$ is the Schur complement of the covariance matrix ${\bf Cov}(x,x)$ for $x := ((\partial_q S)^\top,q^\top)^\top$. As long as a covariance matrix is always positive-semidefinite \cite{kulkarni1999} and a positive-semidefinite matrix has a positive-semidefinite Schur complement \cite{horn2013}, we conclude that ${\bf \Delta} \ge 0$. Therefore, for pure Gaussian states, the uncertainty relation (\ref{urpart}) implies the \hy{RSUR} (\ref{schur}). In particular, (\ref{lur1df}) or (\ref{lur1df2})
implies (\ref{rs1d}) for an arbitrary one-\hy{DF} Gaussian state.

\ %%%%%%%%%%%%%%%%%%%%%%%%%%%%%%%%%%%%%%%%%%%%%%%%%%%%%%%%%%%%%%%%%%%%%%%%%%%%%%%%%%%%%%% 
\paragraph{Absence of Classical Correlations:} %%%%%%%%%%%%%%%%%%%%%%%%%%%%%%%%%%%%%%%%%% 
%%%%%%%%%%%%%%%%%%%%%%%%%%%%%%%%%%%%%%%%%%%%%%%%%%%%%%%%%%%%%%%%%%%%%%%%%%%%%%%%%%%%%%%%% 
The uncertainty relations (\ref{nccorur}) and (\ref{nccorurmix}) 
assert the necessary presence of quantum correlations on any physical system. 
Notwithstanding, there is no bound for the classical correlations, 
hence we shall analyze what happens when 
$\widetilde{\mathbf V}_{\!\text{c}}$ in (\ref{correl}) or (\ref{correlmix}) vanishes. 

For pure states, a sufficient condition to have 
$\widetilde{\mathbf V}_{\!\text{c}} = {\bf 0}_n$ is that $\partial_q S = 0$, 
while for mixed states we need $\partial_q S_k = 0, \forall k$ [see Eq.(\ref{apB3})]. 
In these cases we have $p_\text{c} = 0$ in (\ref{clmom}) and in (\ref{clmomixed}). 
Moreover, since ${\bf V}_{\! qp} = {\bf 0}_n$, the \hy{RSUR} in (\ref{schur}) becomes
\begin{equation}                                                                         \label{urpart2}                                                                                                 
{\bf V} \widetilde{\mathbf V}_{\!\text{nc}} \ge \frac{\hbar^2}{4} \mathsf I_{n} \ , 
\end{equation}   
which is an uncertainty relation including just nonclassical correlations.
 
Typically, this situation occurs for real eigenfunctions of the Hamiltonian operator. Indeed, the unitary evolution of an eigenstate $|\psi_k\rangle$ with eigenvalue $E_k$ is $\exp(-i E_k t/\hbar)|\psi_k\rangle$, while the phase is given by $S_k(q,t) = S_k(q,0) - E_k t/\hbar$.  Consequently, $\partial_q S_k(q,t) = \partial_q S_k(q,0)$. If the phase of the initial state is at most linear in $q$, then $\partial_q S_k(q,t)$ will be a constant vector, and ${\rm \bf Cov}(\partial_q S_k,\partial_q S_k) = {\bf V}_{\! qp} = {\bf 0}_n$.   

For real wave functions, it is convenient to write $\psi_k(q,0) = \Omega_k(q,0)\cos \left[S_k(q,0)/\hbar\right]$. It is well known that eigenfunctions of time-inversion symmetric Hamiltonians are real, consequently, $\partial_q S_k(q,t) = 0, \forall t \in \mathbb R$. Note that this is not the case for the generic Hamiltonian in (\ref{ham}) due to the terms $q \cdot {\mathbf C} p + p \cdot \xi_p$. 

Another example of a state with no classical correlations is the Gaussian state described in (\ref{wfgauss}) where the phase in (\ref{ampph}) has ${\rm Im} ({\bf \Sigma}_{\sf S}) = 0$. This happens when ${\bf b} = {\bf c} = {\bf 0}_n$ or ${\bf a} = {\bf d} = {\bf 0}_n$ in (\ref{smdef}).

From inequality (\ref{urpart2}), one can prove that 
\[
\tfrac{\hbar^2}{4} \, {\rm Tr} \! \left( {\bf V}^{-1}{\bf M} \right)  
\le
{\rm Tr} \left[ \widetilde{\mathbf V}_{\!\text{nc}} {\bf M}\right]\ ,
\]
which is clearly saturated (becomes an equality) for pure Gaussian states since they satisfy (\ref{urpart}). Nevertheless, for mixed states this issue is more involved. 

Let us assume that $\partial_q S_k = 0 \ (\forall k)$ and hence $\delta {\bf V}^{(k)} = {\bf 0}_n$. Furthermore, according to (\ref{correlmix}), there are no classical correlations  either ($\widetilde{\mathbf V}_{\! \text{c}} = {\bf 0}_n$)  since $\partial_q \bar S\big|_{q'=q} = 0$ [see (\ref{apB3})]. 
Therefore, we have $\widetilde{\mathbf V} = \widetilde{\mathbf V}_{\! \text{nc}}$ [see (\ref{correlmix})], 
which becomes the convex combination of the $\widetilde{\mathbf V}_{\! \text{nc}}^{(k)}$'s in (\ref{correlmix2}). This is what happens, for example, when all $|\psi_k\rangle$ in (\ref{spectral}) are eigenstates of the Hamiltonian of the system.  

However, $\delta {\bf V}^{(k)} = {\bf 0}_n\ (\forall k)$ is not a necessary condition for expressing the matrices in (\ref{correlmix}) as convex combinations of the ones in (\ref{correlmix2}). In fact, when all states $|\psi_k\rangle$ in (\ref{spectral}) are such that $\left\langle\partial_q S_k \right\rangle = 0$, then $\delta{\bf V}^{(k)} = {\bf 0}_n \, (\forall k)$, but this does not imply that either one in (\ref{correlmix}) becomes the convex sum of the others in (\ref{correlmix2}).

Note that the same condition for $\delta {\bf V}^{(k)} = {\bf 0}_n$ also implies $\delta \widetilde{\mathbf V}_{\!\text{nc}} = {\bf 0}_n$, see (\ref{apB7}). In this case, (\ref{qpmix}) gives the convex decomposition of $\langle \bar Q (t)\rangle_{\rho}$ in terms of $\left\langle Q_k (t)\right\rangle$.

\ %%%%%%%%%%%%%%%%%%%%%%%%%%%%%%%%%%%%%%%%%%%%%%%%%%%%%%%%%%%%%%%%%%%%%%%%%%%%%%%%%%%%%%% 
\paragraph{Systems with independent $n$-DF:}  %%%%%%%%%%%%%%%%%%%%%%%%%%%%%%%%%%%%%%%%%%%% 
%%%%%%%%%%%%%%%%%%%%%%%%%%%%%%%%%%%%%%%%%%%%%%%%%%%%%%%%%%%%%%%%%%%%%%%%%%%%%%%%%%%%%%%%% 
Let us assume a system with $n$ \hy{DF} that are completely independent from each other, {\it i.e.},  
its evolution is governed by a Hamiltonian given by $H(q,p) = \sum_{i=1}^{n} H_i(q_i,p_i)$ with
\[
H_i(q_i,p_i) := \tfrac{1}{2m_i} p_i^2  + L^{i}q_i p_i + \xi_p^i p_i + U_i(q_i) \ .
\]
Thus, if the system starts in an uncorrelated initial state, it will remain uncorrelated for all times. From (\ref{qp}), the quantum potential also factorizes into
\[
Q(q,t) = \sum_{i=1}^{n} Q_i(q_i,t), \,\,\,
Q_i(q_i,t) := -\frac{\hbar^2}{2m_i\Omega_i} 
              \frac{\partial^2 \Omega_i}{\partial q_i^2}\ .
\]

Now, choosing $n$ functions $T_0^i(q) = \zeta_i q_i \in \mathcal{T}_{\Omega_i}^1$, and following the same reasoning as in Theorem~\ref{thmQ}, one can show that each degree of freedom satisfies an inequality identical to (\ref{lur1df}). Summing all these inequalities, one obtains
\[
\langle Q(t) \rangle = \sum_{i=1}^n \langle Q_i(t) \rangle \ge \frac{\hbar^2}{8} \sum_{i=1}^n \frac{1}{m_i\Delta q^2_i}\ ,
\]
which is different from (\ref{qpurlf}).

%%%%%%%%%%%%%%%%%%%%%%%%%%%%%%%%%%%%%%%%%%%%%%%%%%%%%%%%%%%%%%%%%%%%%%%%%%%%%%%%%%%%%%%%%
%%%%%%%%%%%%%%%%%%%%%%%%%%%%%%%%%%%%%%%%%%%%%%%%%%%%%%%%%%%%%%%%%%%%%%%%%%%%%%%%%%%%%%%%%
\section{Examples} \label{Exm}%%%%%%%%%%%%%%%%%%%%%%%%%%%%%%%%%%%%%%%%%%%%%%%%%%%%%%%%%%%%%%%%%%%%%%
%%%%%%%%%%%%%%%%%%%%%%%%%%%%%%%%%%%%%%%%%%%%%%%%%%%%%%%%%%%%%%%%%%%%%%%%%%%%%%%%%%%%%%%%%
%%%%%%%%%%%%%%%%%%%%%%%%%%%%%%%%%%%%%%%%%%%%%%%%%%%%%%%%%%%%%%%%%%%%%%%%%%%%%%%%%%%%%%%%%
In this section we will exemplify our results with known physical systems. We hope that analyzing concrete examples will help to gain physical insight into our previous conclusions. \\

%%%%%%%%%%%%%%%%%%%%%%%%%%%%%%%%%%%%%%%%%%%%%%%%%%%%%%%%%%%%%%%%%%%%%%%%%%%%%%%%%%%%%%%%%
\paragraph{Harmonic Oscillator Eigenfunctions:}%%%%%%%%%%%%%%%%%%%%%%%%%%%%%%%%%%%%%%%%%%       
%%%%%%%%%%%%%%%%%%%%%%%%%%%%%%%%%%%%%%%%%%%%%%%%%%%%%%%%%%%%%%%%%%%%%%%%%%%%%%%%%%%%%%%%%
Consider the eigenfunctions for a one-\hy{DF} Harmonic Oscillator
\begin{equation}                                                                         \label{exOHeig}
\psi_n(q) = \frac{1}{\sqrt{2^n n!}} 
            \left(\frac{1}{2\pi\Delta q^2_0}\right)^{\!\tfrac{1}{4}} 
            {\rm e}^{-\frac{q^2}{4\Delta q^2_0}} 
            {\rm H}_n \! \left(\frac{q}{\sqrt{2}\Delta q_0}\right)\ , 
\end{equation}
where ${\rm H}_n(x)$ are the Hermite polynomials and $\Delta q_0^2$ is the position variance of the ground state, which is a Gaussian state since ${\rm H}_0(x) = 1$. The variances of these states are given by  
\begin{equation}                                                                         \label{exOHposvar}
\Delta q^2_n =  (2n+1)\Delta q^2_0, \,\,\, 
\Delta p^2_n =  (2n+1)\frac{\hbar^2}{4\Delta q^2_0}\ .
\end{equation}

Considering the harmonic oscillator Hamiltonian as
\[
\hat H = \frac{\nu}{2}(\hat q^2 + \hat p^2)\ ,
\]
and using (\ref{qp}), 
the quantum potential reads
\[
Q_n(q) =   (2n+1) \frac{\hbar^2 \nu}{4 \, \Delta q^2_0} 
                 -\frac{\hbar^2 \nu}{8 \, \Delta q^4_0} q^2,
\]
which is time independent inasmuch as the time evolution does not change the modulus of $\psi_n(q)$.  Since $\langle q \rangle = 0$ for the Harmonic Oscillator eigenfunctions, 
the \hy{MVQP} in (\ref{meanqp}) is readily obtained
\begin{equation}                                                                         \label{OHmeanqp}
\langle Q_n \rangle =  
(2n+1) \Delta q_0^2 \frac{\hbar^2 \nu}{8 \, \Delta q^4_0} = 
                    \frac{\hbar^2 \nu}{8} \frac{\Delta q_n^2}{\Delta q^4_0}\ , 
\end{equation}
where we have used (\ref{exOHposvar}). The function $T_0$ which solves (\ref{extbound}) can be determined using ${\rm H}'_n = 2 x {\rm H}_n-{\rm H}_{n+1}$ \cite{gradshteyn}, such that it is given by
\[
T_\ast(q) =  \alpha + 
          \beta \left[ \frac{q}{\sqrt{2}\Delta q_0} - 
                       \frac{ {\rm H}_{n+1}(x)}{ {\rm H}_n(x) } 
                \right]_{x = \frac{q}{\sqrt{2}\Delta q_0}}\ , 
\]
for $\alpha$ and $\beta$ real constants. As can be seen, the above $T_\ast$ is no longer a linear function, unless $n=0$. From (\ref{auxfunc}), the above $T_\ast(q)$ and $T_1(q)$ are linearly dependent, and thus $\mathcal{L}_{Q}$ is equal to $\langle Q \rangle$ in (\ref{OHmeanqp}). Notwithstanding, the limiting function in (\ref{qpur}) for the linear function is  
\[
{\mathcal L}_Q(\zeta q + \zeta_0) = \frac{\hbar^2 \nu}{8 \Delta q_n^2} 
                                  = \frac{\hbar^2 \nu}{8 (2n + 1) \Delta q_0^2}\ .
\]

Since the linear $T_0(q)$ is a sufficient condition valid for any state, see Sec.\ref{scfe}{\color{Blue}a},
the function above constitutes a bound for \hy{MVQP}. Therefore the uncertainty relation for the wave functions (\ref{exOHeig}) is written as (\ref{lur1df}) with $\Delta q = \Delta q_n$. Furthermore, the comparison of the above bound with (\ref{qpurlf}) shows that $\langle Q \rangle = {\mathcal L}_Q(a q + b)$ only for $n=0$, which is a pure Gaussian state. This shows that none of the Harmonic Oscillator eigenfunctions with $n>0$ saturates the linear uncertainty relation, as expected.

As eigenstates, the classical correlations are null and the one-\hy{DF} version of (\ref{urpart2}) applies. Using (\ref{meanqpcorr}), relation (\ref{urpart2}) becomes 
\[
\Delta q^2_n \langle Q_n \rangle = (2n+1)^2 \frac{\hbar^2\nu}{8} \ge \frac{\hbar^2\nu}{8}\ ,
\]
which is exactly (\ref{lur1df}). 

\ %%%%%%%%%%%%%%%%%%%%%%%%%%%%%%%%%%%%%%%%%%%%%%%%%%%%%%%%%%%%%%%%%%%%%%%%%%%%%%%%%%%%%%%
\paragraph{Thermal State:}%%%%%%%%%%%%%%%%%%%%%%%%%%%%%%%%%%%%%%%%%%%%%%%%%%%%%%%%%%%%%%% 
%%%%%%%%%%%%%%%%%%%%%%%%%%%%%%%%%%%%%%%%%%%%%%%%%%%%%%%%%%%%%%%%%%%%%%%%%%%%%%%%%%%%%%%%%
As an example of a mixed state, 
let us consider the thermal state of the harmonic oscillator. 
The density operator is written as (\ref{spectral}), with 
\[
\omega_k = \frac{{\rm e}^{- \hbar \beta \nu k}}{\bar n + 1}\quad  \mbox{and} \quad 
\bar n := ({\rm e}^{\hbar \beta \nu } - 1)^{-1}\ .
\]
The parameter $\beta$ is the inverse of the temperature, $|\psi_k\rangle$ are the harmonic oscillator eigenstates with eigenfunctions given by (\ref{exOHeig}) and the sum in (\ref{spectral}) ranges in $k =0,1,2,...,\infty$. As discussed in Sec.\ref{pmu}, the classical correlations for this state vanish, since the eigenstates of the spectral decomposition are also eigenstates of the Hamiltonian of the system. 

The position and momentum variances for the thermal state are obtained as the convex sum of the ones for the Harmonic oscillator eigenfunctions (\ref{exOHposvar}), namely
\begin{align*}
\Delta q^2 &= {\rm cotanh}\left( \tfrac{1}{2}\hbar \beta \nu \right) \Delta q_0^2\ , \\ 
\ \Delta p^2 &= {\rm cotanh}\left( \tfrac{1}{2}\hbar \beta \nu \right) \Delta p_0^2\ .
\end{align*}

Since there are no classical correlations for this system, the above $\Delta p^2$ is due solely to the nonclassical part of the momentum. In addition, from (\ref{qpmix}) with $\delta \widetilde{\mathbf V}_{\!\text{nc}} = {\bf 0}_n$, the \hy{MVQP} in (\ref{meanqpmix}) becomes the convex sum of the ones in (\ref{OHmeanqp})
\[
\langle \bar Q \rangle_\rho = 
\frac{\hbar^2 \nu}{8 \Delta q_0^2}\,
{\rm cotanh}\left( \tfrac{1}{2}\hbar \beta \nu \right) = 
\frac{\nu}{2} \Delta p^2\ .
\] 

Note that the \hy{QP} is a monotonically decreasing function of $\beta$, which shows that the quantum correlations are erased for lower temperatures. Nevertheless, the limit $\lim_{\beta \to \infty} \langle \bar Q \rangle_\rho = \frac{\nu}{2} \Delta p_0^2$ is finite; hence, even for vanishing temperatures the correlations in momentum remain. In general, correlations, such as entanglement, are expected to vanish for high temperatures \cite{DBLP:journals/qic/AndersW08}. However, there exist also correlations, such as quantum discord \cite{PhysRevLett.105.020503} that increase with the temperature, similarly to the correlations of momenta described above.

\ %%%%%%%%%%%%%%%%%%%%%%%%%%%%%%%%%%%%%%%%%%%%%%%%%%%%%%%%%%%%%%%%%%%%%%%%%%%%%%%%%%%%%%%
\paragraph{Coherent and Squeezed States:}%%%%%%%%%%%%%%%%%%%%%%%%%%%%%%%%%%%%%%%%%%%%%%%%
%%%%%%%%%%%%%%%%%%%%%%%%%%%%%%%%%%%%%%%%%%%%%%%%%%%%%%%%%%%%%%%%%%%%%%%%%%%%%%%%%%%%%%%%%  
Let us consider a system with $n$ \hy{DF} described initially by a Gaussian state, see (\ref{wfgauss}), with symplectic matrix given by
\begin{equation}                                                                         \label{excvgauss}
{\mathsf S} = {\bf a} \oplus {\bf d}\quad , \quad 
{\bf a}={\bf d}^{-1}, \,\,\, {\bf a} = {\rm Diag}(a_{11},...,a_{nn})\ ,  
\end{equation}
which means that the system is uncorrelated and described by the covariance matrix
\[
{\mathbf V} = {\rm Diag}(\Delta q_{1}^2,...,\Delta q_{n}^2)\ , 
\]
with $\Delta q_i = \sqrt{\tfrac{\hbar}{2}}\, |a_{ii}|$ for $(i=1,...,n)$ and $a_{ii}$ are the squeezing parameters of the state. The evolution is given by the Hamiltonian of $n$ noninteracting harmonic oscillators, 
\begin{equation}                                                                         \label{exohham}
\hat H = \sum_{i=1}^n \frac{\nu_i}{2}(\hat q_i^2 + \hat p_i^2)\ ,
\end{equation}
which can be brought to the form in (\ref{ham}) 
with potential energy in (\ref{quadpot})  
by setting ${\bf M} = {\rm Diag}(\nu_1,...,\nu_n)$. Using (\ref{sympdyn}), the symplectic matrix generated by this Hamiltonian reads
\[
{\mathsf S}_t = 
\left( \!\! \begin{array}{rc}
            \cos({\bf M} t)  & \sin({\bf M} t)   \\
           -\sin({\bf M} t)  & \cos({\bf M} t)   
           \end{array}
\right), \,\,\, 
{\bf M}_{ij} = \nu_i \delta_{ij}\ ,
\]
and the evolved state, {\it cf.}~(\ref{simpev}), will also be a Gaussian state with covariance matrix
\begin{equation}                                                                         \label{excvgauss3}
{\bf V}_t  =  \frac{\hbar}{2} 
              [ \cos^2({\bf M} t) {\bf aa}^\top  + 
                \sin^2({\bf M} t)  {\bf dd}^\top]\ ;
\end{equation}
if the initial mean value of position and momentum are, respectively, $\eta_q$ and $\eta_p$, the position mean value vector reads
\begin{equation}                                                                         \label{exmvvec}
{\eta_q}(t) = \cos({\bf M} t) \eta_q + \sin({\bf M} t)\eta_p\ .
\end{equation}

Since the initial state is uncorrelated and the dynamics is described by a non-interacting Hamiltonian, the \hy{MVQP} in (\ref{meangaussqp}), as well as the \hy{QP}, becomes a sum of terms each for one degree of freedom, {\it i.e.}
\begin{equation}                                                                         \label{exglobqp}
\langle Q(t)\rangle   = \sum_{i=1}^n \langle Q_i(t)\rangle\ , 
\end{equation}
where
\begin{equation}                                                                         \label{exindqp}
\langle Q_i(t)\rangle = 
\frac{\hbar^2 \nu_i}{8\Delta q_i^2}  
\left[\cos^2(\nu_i t) + a_{ii}^{-4} \sin^2(\nu_i t)\right]^{-1}\ .
\end{equation}

Noting that 
\[
\Delta q_i(t)  = \Delta q_i 
\left[\cos^2(\nu_i t) + a_{ii}^{-4} \sin^2(\nu_i t)\right]^{-1/2}\ ,
\]
each of the individual quantum potentials saturates the uncertainty relation as expected by (\ref{qp1dfur}).  By the other side, the full \hy{MVQP} (\ref{exglobqp}) satisfies (\ref{qpurlf}), which becomes
\[
\langle Q(t)\rangle \ge 
\frac{\hbar^2 }{8} \, 
{\rm max} \left\{ \frac{\nu_i}{ [\Delta q_i(t)]^2}; \, {i=1,...,n} \right\}\ . 
\]

Let us now consider a $n$-\hy{DF} coherent state, which can be obtained by setting $a_{ii} = 1, \forall i$ in (\ref{excvgauss}). This means that its wave function is obtained from (\ref{wfgauss}) by choosing $\mathsf S = \mathsf I_{2n}$ in (\ref{smdef}). It is well known~\cite{Sakurai1994,Cohen1977} that the Hamiltonian (\ref{exohham}) preserves the coherent character of the state.

The \hy{QP} in (\ref{meangaussqp}) will be the sum of the individual \hy{QP}s 
\[
Q_i(q_i,t) = \frac{\hbar^2 \nu_i }{4 \Delta q_i^2} 
       - \frac{\hbar^2 \nu_i }{8\Delta q_i^4}(q_i - {\eta_q}_i(t))^2\ ,
\]
and will be time dependent through the mean value vector in (\ref{exmvvec}). Notwithstanding, the \hy{MVQP} will be time independent since it depends only on the covariance matrix, which is constant, as can be seen by setting ${\bf a} = {\mathsf I}_n$ in (\ref{excvgauss3}). Thus, using (\ref{exindqp}), we have $\langle Q_i\rangle = \hbar \nu_i/4$.

\ %%%%%%%%%%%%%%%%%%%%%%%%%%%%%%%%%%%%%%%%%%%%%%%%%%%%%%%%%%%%%%%%%%%%%%%%%%%%%%%%%%%%%%%
\paragraph{Non Linear Functions for the Bound:}%%%%%%%%%%%%%%%%%%%%%%%%%%%%%%%%%%%%%%%%%%%%%%%%%%%%%
%%%%%%%%%%%%%%%%%%%%%%%%%%%%%%%%%%%%%%%%%%%%%%%%%%%%%%%%%%%%%%%%%%%%%%%%%%%%%%%%%%%%%%%%%  
It is interesting to see what changes when we choose different functions $T_0(q)$ for the bound in (\ref{qpur}). We will also consider a one-\hy{DF} Gaussian state evolving subjected to the Hamiltonian (\ref{ham}) with a quadratic potential. Let us ignore the solution that extremizes the bound (\ref{qpur}) and choose a power law function of the form
\[
T_0(q) = (q - \eta_q)^k, \,\,\, k \in {\mathbb N}.   
\]
A known result about the {\it centered moments} of a Gaussian distribution \cite{papoulis} is that its mean reads
\[
\langle (q - \eta_q)^k \rangle = 
\left\{ \begin{array}{l}
        0 \,\,\, \text{if} \,\,\, k \,\,\, \text{odd}\ , \\
        \Delta q^k (k-1)!! \,\,\, \text{if} \,\,\, k \,\,\, \text{even}\ . 
                                       \end{array}  \right.
\]

Inserting $T_0(q)$ above in (\ref{qpur}) and using the above moments, one can show that 
\[
{\mathcal L}_{Q}((q-\eta_q)^k,t) = 
C_k \, {\mathcal L}_{Q}(q-\eta_q,t)\ ,
\]
where
\[
C_k := 
\left\{ \begin{array}{cl}
        \frac{(k!!)^2}{(2k-1)!!} &\,\,\, \text{if} \,\,\, k \,\, \text{odd}\ , \\
        0 &\,\,\, \text{if} \,\,\, k \,\, \text{even}\ . 
        \end{array}  \right.
\]
The coefficients $C_k$ satisfy the following properties
\[
\begin{split}
&\text{(i)}\,\, C_1 = 1\ , \\ 
&\text{(ii)}\,\, C_k \ge 0, \forall k \in \mathbb N\ , \\ 
&\text{(iii)}\,\, C_{2k+1} > C_{2k+3}, \forall k \in \mathbb N\ . 
\end{split}
\]
Properties (i) and (ii) are straightforward, while (iii) can be proved by induction. These properties show that ${\mathcal L}_{Q}((q-\eta_q)^k,t) \le {\mathcal L}_{Q}((q-\eta_q),t), 
\forall k \in \mathbb N$. This agrees with the fact that the linear bound in (\ref{qpurlf}) is the greatest bound for Gaussian states.

\ %%%%%%%%%%%%%%%%%%%%%%%%%%%%%%%%%%%%%%%%%%%%%%%%%%%%%%%%%%%%%%%%%%%%%%%%%%%%%%%%%%%%
\paragraph{Inverted Oscillator:}%%%%%%%%%%%%%%%%%%%%%%%%%%%%%%%%%%%%%%%%%%%%%%%%%%%%%%
%%%%%%%%%%%%%%%%%%%%%%%%%%%%%%%%%%%%%%%%%%%%%%%%%%%%%%%%%%%%%%%%%%%%%%%%%%%%%%%%%%%%%%
Consider the one-\hy{DF} Hamiltonian 
\[
\hat H = \frac{\nu}{2}(\hat p^2 - \hat q^2)\ ,  
\]
which describes a scattering interaction through a parabolic barrier. An initial coherent state evolves into a kind of squeezed state with wave function (\ref{wfgauss}) determined by 
\[
{\mathsf S}_t = 
\left(  \begin{array}{rc}
           \cosh(\nu t)  & \sinh(\nu t)   \\
           \sinh(\nu t)  & \cosh(\nu t)   
        \end{array} 
\right), 
\]
which implies ${\bf V} = {\Delta q^2} = \frac{\hbar}{2} \cosh(2\nu t)$. The quantum potential given by (\ref{meangaussqp}) reads $\langle Q(t) \rangle = \hbar\nu/[4 \cosh (2\nu t)]$. 

As $t$ increases, the dispersion on the position increases, while the mean value of the QP becomes smaller in order to maintain the relation $\langle Q(t)\rangle \Delta q^2 = \hbar^2 /(8m)$ intact. However, this saturation does not happen for the Heisenberg uncertainty relation. The dispersion on the momentum can be calculated directly from the wave function and gives $\Delta p^2 = \hbar \cosh(2\nu t)/2$. Thus, we have $\Delta q^2\Delta p^2 > \hbar^2/4$ for all $t>0$.

\ %%%%%%%%%%%%%%%%%%%%%%%%%%%%%%%%%%%%%%%%%%%%%%%%%%%%%%%%%%%%%%%%%%%%%%%%%%%%%%%%%%%%%%%
\paragraph{P\"oschl-Teller Potential:}%%%%%%%%%%%%%%%%%%%%%%%%%%%%%%%%%%%%%%%%%%%%%%%%%%%
%%%%%%%%%%%%%%%%%%%%%%%%%%%%%%%%%%%%%%%%%%%%%%%%%%%%%%%%%%%%%%%%%%%%%%%%%%%%%%%%%%%%%%%%%  
The system described by the Hamiltonian 
\[
\hat H = \frac{\hat p^2}{2m} - \frac{\lambda(\lambda+1)}{2} {\rm sech}(\hat q) \ ,
\]
constitutes one of the few examples of an analytically solvable problem in quantum mechanics \cite{poschl1933}. The eigenfunctions and the Hamiltonian eigenvalues for this potential are, respectively, given by  
\[
\psi_{\lambda}^\mu(q) = 
                        \sqrt{ \frac { \mu (\lambda - \mu)! }
                                     {     (\lambda + \mu)! } } 
                        \, {\rm P}_\lambda^\mu(\tanh(q))\ , 
\,\,\, 
E_\mu = -\frac{\hbar^2 \mu}{2m}\ ,
\]
where $P_\lambda^\mu(x)$ are the Legendre associated polynomials \cite{gradshteyn}, $\lambda \in \mathbb N$, and  $\mu = 1,2,...,\lambda-1,\lambda$. Inserting the wave function in (\ref{qp}) and using the identity \cite{gradshteyn}
\[
{\rm P}_{\lambda+1}^{\mu }(x) = \frac{(2 \lambda +1)}{(\lambda -\mu +1)} 
                                 x {\rm P}_{\lambda}^{\mu }(x) 
                              - \frac{(\lambda +\mu)}{(\lambda -\mu +1)} 
                                  {\rm P}_{\lambda-1}^{\mu }(x)\ ,
\]
the \hy{QP} and its \hy{MVQP} for the P\"oschl-Teller potential read
\begin{align*}
Q_{\lambda}^{\mu}(q) &= - E_\mu + \frac{\hbar^2}{2m} \lambda  (\lambda +1) {\rm sech}^2(q)\ ,\\
\langle Q \rangle &= - E_\mu + \frac{\hbar^2}{2m} \frac{2 \mu \lambda (\lambda +1)}{2 \lambda +1}\ .
\end{align*}

From (\ref{extbound}), the function that extremizes the inequality is 
\[
T_\ast(q) = 2 (\lambda +1) {\tanh}(q) - 
2 \alpha (\lambda -\mu +1) 
\frac{{\rm P}_{\lambda +1}^{\mu }(\tanh(q))}
{{\rm P}_{\lambda }^{\mu }(\tanh(q))}\ , 
\]
for a real constant $\alpha$. In order to study the behavior of the bound function (\ref{qpur}), we will choose $\lambda = \mu$, which corresponds to the highest excited state of the P\"oschl-Teller potential for a given $\lambda$.

According to (\ref{auxfunc}), the saturation of the Cauchy-Schwartz inequality (see Sec.\ref{pmu}) happens when $T_\ast(q)$ becomes proportional to $T_1(q) = {\tanh}(q)$. As a fact, using the identity \cite{gradshteyn} ${\rm P}_{\mu +1}^{\mu }(x) = (2\mu + 1) x {\rm P}_{\mu +1}^{\mu}(x) $ in the above function $T_\ast(q)$ it follows that $\left\langle Q \right\rangle = \mathcal L_{Q}({\tanh}(q))$. 

\begin{figure}[h]
\includegraphics[width=7.5cm, trim=0 25 0 0]{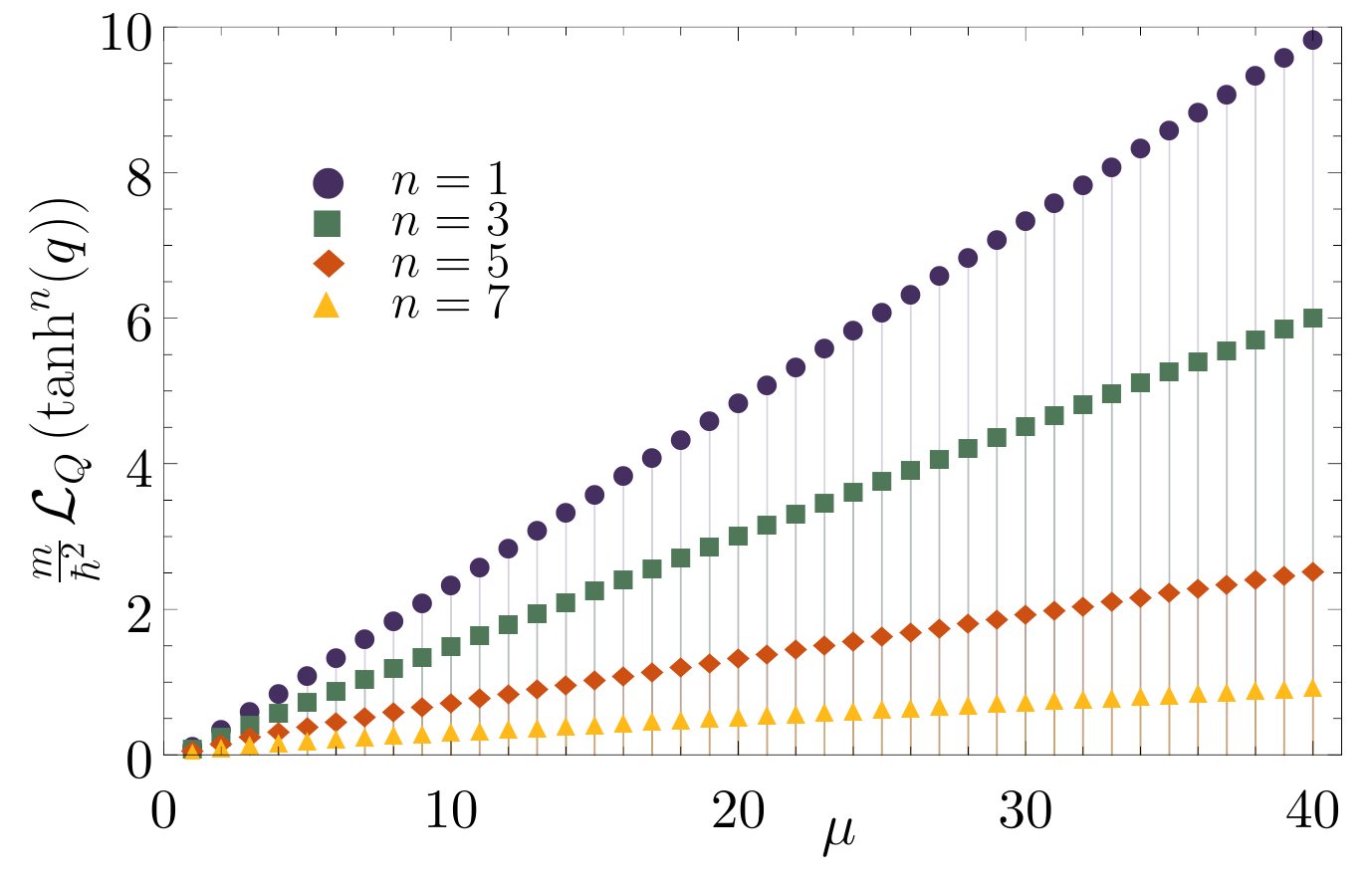}
\caption{ (Color Online) 
          Bound $\mathcal L_{Q}(T_0(q))$ of the quantum potential 
          associated to the eigenstates of the P\"oschl-Teller 
          potential for the function 
          $T_0(q) = \tanh^n(q)$. The case $n=1$ 
          corresponds to the mean value of the quantum potential, 
          {\it i.e.}, 
          $\mathcal L_{Q}(\tanh(q)) = \left\langle Q \right\rangle$.}                    \label{fig1}
\end{figure}

The uncertainty relation (\ref{qpur}) can be (analytically) determined by considering functions of the form $T_0(q) = {\tanh}^n(q)$ with $n \in \mathbb N$. In this case, we find
\[
\begin{split}
\left\langle T_0 \right\rangle &= \sqrt{ \frac{ 1 + (-1)^n  }{2 \pi} } \, 
\frac{ \Gamma \! \left( \mu \!+\! \frac{1}{2}\right)
       \Gamma \! \left(\frac{n+1}{2} \right)           }
     { \Gamma \! \left(\frac{n+1}{2}\! + \!\mu \right) } \ ,                                \\
\left\langle T_0^2 \right\rangle &= 
\frac{1}{\sqrt{\pi }} 
\frac{ \Gamma \! \left( \! \mu \! + \! \frac{1}{2} \right) 
       \Gamma \! \left( \frac{n}{2} \! + \! 1 \right)           }
     { \Gamma \! \left( \frac{n}{2} \! + \! \mu \! + \! 1 \right) } \ ,                                \\
\left\langle{\partial_q T_0}\right\rangle  &=                   
\left[1-(-1)^n\right] \mu \left\langle T_0^2 \right\rangle\ .  
\end{split}
\]
Therefore, the bound function vanishes if $n$ is even, $\mathcal{L}_Q({\tanh}^{n}(q),t) = 0$, and gives 
\[
\mathcal{L}_Q({\tanh}^{n}(q),t) = 
\frac{ \mu^2 \hbar^2 }{2 m}
\frac{ \Gamma \! \left( \mu \! + \!\frac{1}{2} \right) 
       \Gamma \! \left( n \!+\! \mu \!+\! \frac{1}{2} \right) 
       \Gamma \! \left( \frac{n}{2} \!+ \!1 \right)^2     }
     { \sqrt{\pi} \, 
       \Gamma \! \left( n \!+ \!\frac{1}{2} \right) 
       \Gamma \! \left( \frac{n}{2}\! + \! \mu \! + \! 1 \right)^2 }, 
\]
for $n$ odd. Figure~\ref{fig1} shows this bound for some values of $n$. 

%%%%%%%%%%%%%%%%%%%%%%%%%%%%%%%%%%%%%%%%%%%%%%%%%%%%%%%%%%%%%%%%%%%%%
%%%%%%%%%%%%%%%%%%%%%%%%%%%%%%%%%%%%%%%%%%%%%%%%%%%%%%%%%%%%%%%%%%%%%
%%%%%%%%%%%%%%%%%%%%%%%%%%%%%%%%%%%%%%%%%%%%%%%%%%%%%%%%%%%%%%%%%%%%%
%%%%%%%%%%%%%%%%%%%%%%%%%%%%%%%%%%%%%%%%%%%%%%%%%%%%%%%%%%%%%%%%%%%%%
%%%%%%%%%%%%%%%%%%%%%%%%%%%%%%%%%%%%%%%%%%%%%%%%%%%%%%%%%%%%%%%%%%%%%
%%%%%%%%%%%%%%%%%%%%%%%%%%%%%%%%%%%%%%%%%%%%%%%%%%%%%%%%%%%%%%%%%%%%%

Let us now consider the uncertainty relation (\ref{lur1df}) for a linear function $T_0(q) = \zeta q + \zeta_0$. In accordance with (\ref{lur1df}), we need to determine the position variance, which is given by a generalized hypergeometric function $_r{\rm F}_s[(a_r);(b_s);z]$ \cite{gradshteyn}. Indeed, we have  
\[
\begin{split}
\Delta q^2 & = \int^{+\infty}_{-\infty} \!\!\!\! [{\psi}_{\mu}^\mu(q)]^2 q^2 dq                     
             = \frac{ (2\mu -1)!! }{ (2\mu -2)!! } 
               \int^{\infty}_{0}\!\!\!\!\! {\rm sech}^{2\mu}(q) \, q^2 dq  
                    \\
&= \frac{ 2^{ \mu - 1 } (2 \mu - 1)!! }{\mu^2 \,\mu!} \,
_4{\rm F}_3 \!\left[ \begin{array}{c}
                   \mu \hspace{0.4cm} \mu \hspace{0.4cm} \mu \hspace{0.4cm} 2\mu \\
                   \mu+1 \; \mu+1 \; \mu+1  
                   \end{array} ; -1 \right]\ .  
\end{split}
\]

\begin{figure}[h]
\includegraphics[width=7.5cm, trim=0 25 0 0]{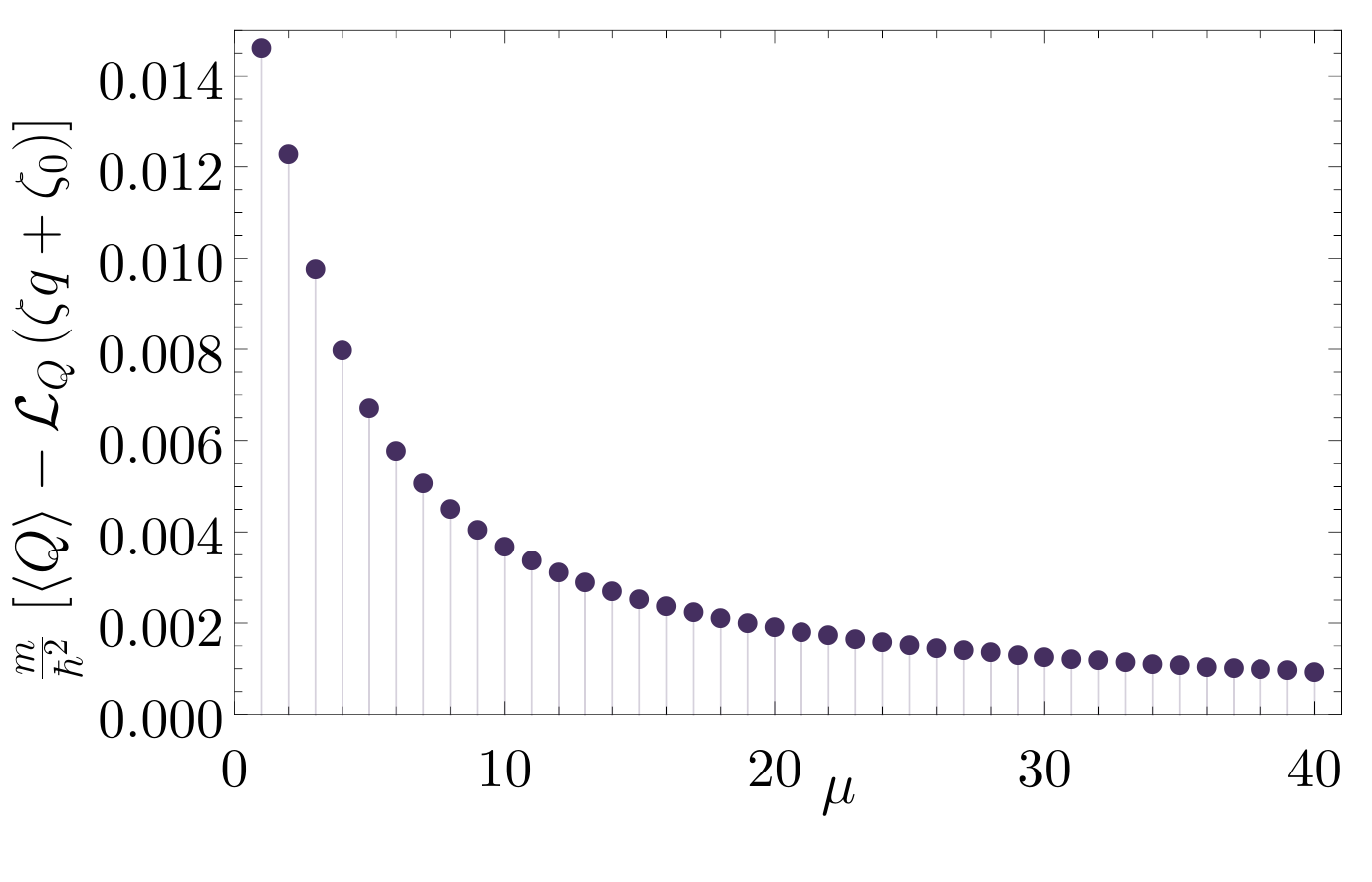}
\caption{ (Color Online) 
          Difference between the mean value of the quantum potential
          and the bound $\mathcal L_{Q}(\zeta q + \zeta_0)$ for 
          the eigenstates of the P\"oschl-Teller potential. 
          The mean value $\left\langle Q \right\rangle$ is the 
          curve $n=1$ in Fig.\ref{fig1}.}                                                \label{fig2}
\end{figure}

In Fig.\ref{fig2}, we compare the bound function with the \hy{MVQP} (case $n=1$ in Fig.\ref{fig1}). Note that the uncertainty relation for the linear function (\ref{lur1df}) gives a stronger constraint than the next function in Fig.\ref{fig1}, {\it i.e.}, the case $n=3$.

%%%%%%%%%%%%%%%%%%%%%%%%%%%%%%%%%%%%%%%%%%%%%%%%%%%%%%%%%%%%%%%%%%%%%
%%%%%%%%%%%%%%%%%%%%%%%%%%%%%%%%%%%%%%%%%%%%%%%%%%%%%%%%%%%%%%%%%%%%%
%%%%%%%%%%%%%%%%%%%%%%%%%%%%%%%%%%%%%%%%%%%%%%%%%%%%%%%%%%%%%%%%%%%%%
%%%%%%%%%%%%%%%%%%%%%%%%%%%%%%%%%%%%%%%%%%%%%%%%%%%%%%%%%%%%%%%%%%%%%
%%%%%%%%%%%%%%%%%%%%%%%%%%%%%%%%%%%%%%%%%%%%%%%%%%%%%%%%%%%%%%%%%%%%%
%%%%%%%%%%%%%%%%%%%%%%%%%%%%%%%%%%%%%%%%%%%%%%%%%%%%%%%%%%%%%%%%%%%%%

As a final remark, note that the wave function for $\lambda=\mu$ corresponds to the most excited states. As $\mu$ increases, the behavior of the system approaches a plane wave and $Q(q,t) \to 0$ for $q\to \pm \infty$. Numerical tests up to $n \sim 10^3$ show that the monotonic decrease observed in Fig.\ref{fig2} is persistent. Thus the quantum potential approaches the linear bound function in this limit, even though the wave function of a free particle does not belong to the set $L^2(\Omega^2)$, and the Cauchy-Schwartz inequality is not applicable.

%%%%%%%%%%%%%%%%%%%%%%%%%%%%%%%%%%%%%%%%%%%%%%%%%%%%%%%%%%%%%%%%%%%%%%%%%%%%%%%%%%%%%%%%%
%%%%%%%%%%%%%%%%%%%%%%%%%%%%%%%%%%%%%%%%%%%%%%%%%%%%%%%%%%%%%%%%%%%%%%%%%%%%%%%%%%%%%%%%%
\section{Conclusions and Final Comments} \label{Concl} %%%%%%%%%%%%%%%%%%%%%%%%%%%%%%%%%%
%%%%%%%%%%%%%%%%%%%%%%%%%%%%%%%%%%%%%%%%%%%%%%%%%%%%%%%%%%%%%%%%%%%%%%%%%%%%%%%%%%%%%%%%%
%%%%%%%%%%%%%%%%%%%%%%%%%%%%%%%%%%%%%%%%%%%%%%%%%%%%%%%%%%%%%%%%%%%%%%%%%%%%%%%%%%%%%%%%%

The debate on interpretation of quantum mechanics has been centered on the properties of the quantum potential and rarely makes any connection between the \hy{QP} and the uncertainty relations. Instead of focusing on the quantum potential, in the present work we analyzed the properties and physical meaning of its mean value. 

In sec.~\ref{GURPS}, we showed that the \hy{MVQP} satisfies an inequality for an arbitrary scalar function $T_0\in L^2\left(\Omega^2\right)$ and, by suitably choosing this function, the \hy{MVQP} is always positive and bounded from below. Furthermore, we derived a generalized uncertainty relation that is stronger than the Robertson-Schr\"odinger inequality.

The physical meaning of the \hy{MVQP} is that it is related to the nonclassical part of the momentum covariance matrix. Decomposing it as $\widetilde{\mathbf V}=\widetilde{\mathbf V}_{\! \text{c}}+\widetilde{\mathbf V}_{\! \text{nc}}$, where $\widetilde{\mathbf V}_{\! \text{c}}$ is exactly the momentum covariance matrix of the classical Hamilton-Jacobi formalism, {\it i.e.} $\widetilde{\mathbf V}_{\! \text{c}}  =   {\mathbf{Cov}}\! \left(\partial_q  S,\partial_q  S \right)$, the nonclassical part identifies with the \hy{MVQP}, namely, $\widetilde{\mathbf V}_{\!\text{nc}}  :=  \left\langle - \frac{\hbar^2}{\Omega}\partial^2_{qq} \Omega \right\rangle$. Thus, the bound on \hy{MVQP} implies that any quantum system has a minimum of quantum momentum correlation. While classical systems can have zero momenta correlations, quantum systems are always correlated.

The results obtained primarily for pure states are then generalized for density matrices describing mixed states. Using a spectral decomposition $\hat \rho = \sum_k \omega_k | \psi_k \rangle \! \langle \psi_k|$,  where $| \psi_k \rangle$ are eigenstates, neither $\widetilde{\mathbf V}_{\! \text{c}}$ nor $\widetilde{\mathbf V}_{\! \text{nc}}$ can be decomposed exclusively as a convex combination of $\widetilde{\mathbf V}_{\! \text{c}}^{(k)}$ nor $\widetilde{\mathbf V}_{\! \text{nc}}^{(k)}$, respectively. Notwithstanding, $\widetilde{\mathbf V}_{\! \text{nc}}$ can still be written as a convex sum, namely, $\widetilde{\mathbf V}_{\! \text{nc}}=\sum_k \omega_k 
\widetilde{\mathbf V}_{\! \text{nc}}^{(k)}+\delta \widetilde{\mathbf V}_{\! \text{nc}}$ where the latter term is a symmetric positive-semidefinite matrix. As a consequence, the \hy{MVQP} defined in~\eqref{meanqpmix} for mixed states is always greater than or equal to the sum of the \hy{MVQP} for each pure state in the spectral decomposition. As a corollary, the \hy{MVQP} for mixed states also has a positive lower bound.

The identification of the \hy{MVQP} with the nonclassical part of the momentum covariance matrix allow us to interpret the semiclassical limit in an adequate manner, which might give new insights for this regime. Indeed, in reference \cite{maia2008}, the WBK propagation becomes a good description for the dynamics of the system when a dynamical stretching of the initial (non-WKB) Gaussian state along an unstable manifold of the classical dynamics is performed. On the lines developed here, this stretching is responsible for the vanishing of the quantum correlations. Thus, our studies might open up new connections in the influence of the quantum potential on the semiclassical propagation.

%%%%%%%%%%%%%%%%%%%%%%%%%%%%%%%%%%%%%%%%%%%%%%%%%%%%%%%%%%%%%%%%%%%%%%%%%%%%%%%%%%%%%%%%
%%%%%%%%%%%%%%%%%%%%%%%%%%%%%%%%%%%%%%%%%%%%%%%%%%%%%%%%%%%%%%%%%%%%%%%%%%%%%%%%%%%%%%%%
\section*{Acknowledgements} %%%%%%%%%%%%%%%%%%%%%%%%%%%%%%%%%%%%%%%%%%%%%%%%%%%%%%%%%%%%
%%%%%%%%%%%%%%%%%%%%%%%%%%%%%%%%%%%%%%%%%%%%%%%%%%%%%%%%%%%%%%%%%%%%%%%%%%%%%%%%%%%%%%%%
%%%%%%%%%%%%%%%%%%%%%%%%%%%%%%%%%%%%%%%%%%%%%%%%%%%%%%%%%%%%%%%%%%%%%%%%%%%%%%%%%%%%%%%%
FN thanks M.M. Taddei for an insightful discussion and Prof. M.E. Vares for her kindly attention on mathematical aspects of this work. FN is a member of the Brazilian National Institute of Science and Technology of Quantum Information [CNPq INCT-IQ (465469/2014-0)] and also acknowledges PROCAD2013. FTF would like to thank and acknowledge financial support from the National Scientific and Technological Research Council (CNPq, Brazil).
%%%%%%%%%%%%%%%%%%%%%%%%%%%%%%%%%%%%%%%%%%%%%%%%%%%%%%%%%%%%%%%%%%%%%%%%%%%%%%%%%%%%%%%%%

%%%%%%%%%%%%%%%%%%%%%%%%%%%%%%%%%%%%%%%%%%%%%%%%%%%%%%%%%%%%%%%%%%%%%%%%%%%%%%%%%%%%%%%%%
%%%%%%%%%%%%%%%%%%%%%%%%%%%%%%%%%%%%%%%%%%%%%%%%%%%%%%%%%%%%%%%%%%%%%%%%%%%%%%%%%%%%%%%%%
\section*{Appendices}  \appendix             %%%%%%%%%%%%%%%%%%%%%%%%%%%%%%%%%%%%%%%%%%%%
%%%%%%%%%%%%%%%%%%%%%%%%%%%%%%%%%%%%%%%%%%%%%%%%%%%%%%%%%%%%%%%%%%%%%%%%%%%%%%%%%%%%%%%%%
%%%%%%%%%%%%%%%%%%%%%%%%%%%%%%%%%%%%%%%%%%%%%%%%%%%%%%%%%%%%%%%%%%%%%%%%%%%%%%%%%%%%%%%%%
\section{Pure Gaussian States}   %\label{pgs}   %%%%%%%%%%%%%%%%%%%%%%%%%%%%%%%%%%%%%%%%%
%%%%%%%%%%%%%%%%%%%%%%%%%%%%%%%%%%%%%%%%%%%%%%%%%%%%%%%%%%%%%%%%%%%%%%%%%%%%%%%%%%%%%%%%%
In this appendix we summarize some information about pure Gaussian states of a system with $n$ degrees of freedom and their symplectic evolution.

It is well known that a quadratic Hamiltonian generates a symplectic evolution and that this preserves the Gaussian character of an initial Gaussian state \cite{littlejohn1986}. The most generic quadratic Hamiltonian can be constructed from the Hamiltonian in (\ref{ham}) with 
\begin{equation}                                                                         \label{quadpot}
U(q) = \tfrac{1}{2} q\cdot {\bf L} q + \xi_q\cdot q + H_0  
\end{equation}
where ${\bf L} = {\bf L}^\top$ is a $n\times n$ symmetric real matrix, 
$\xi_q \in \mathbb R^n$ is a column vector, and $H_0$ is a real constant. Such a Hamiltonian is the generator of the uniparametric symplectic subgroup constituted by $\mathsf S_t$ such that
\begin{equation}                                                                         \label{sympdyn}
{\mathsf S_t} := {\rm e}^{\mathsf J {\bf H} t}, \,\,\,
{\mathsf J} := \left(\!\begin{array}{cc}
                0_n & \mathsf I_n \\
                - \mathsf I_n & 0_n
               \end{array}\!\right), \,\,\,
{\bf H} := \left(\!\begin{array}{lc}
                {\bf L} & {\bf C} \\
                {\bf C}^\top &{\bf M}
               \end{array}\!\right).               
\end{equation}

A $2n\times 2n$ real matrix $\mathsf S$ is said to be symplectic if $\mathsf S\!^\top \!\!\mathsf J \mathsf S = \mathsf S \mathsf J \mathsf S\!^\top = \mathsf J$, which is the case of $\mathsf S_t$ in (\ref{sympdyn}). These matrices can be partitioned into $n \times n$ blocks ${\bf a},{\bf b},{\bf c},{\bf d}$, as follows 
\begin{equation}                                                                         \label{smdef}
\mathsf S = \!
\left(\!
\begin{array}{cc}
\mathbf{a} & \mathbf{b}\\
\mathbf{c} & \mathbf{d}
\end{array}\!
\right)\! 
\, {\rm with}
\left\{ \!\!
\begin{array}{ll}
\mathbf{ad\!^{\top}} - \mathbf{bc\!^{\top}} = \mathsf{I}_n, &
\!\! \mathbf{a\!^{\top}\!c =c\!^{\top}\!a}, \\
\mathbf{a\!^{\top}\!d} - \mathbf{c\!^{\top}\!b} = \mathsf{I}_n, & 
\!\! \mathbf{ab\!^{\top} = ba\!^{\top}}, \\ 
\mathbf{cd\!^{\top} = dc\!^{\top}},  &
\!\! \mathbf{b\!^{\top}\!d = d\!^{\top}\!b} .                           
\end{array}
\right.                                                              
\end{equation}
The constraints over the blocks came from the symplectic nature of $\mathsf S$ \cite{littlejohn1986}.

The wave function of a generic pure Gaussian state of $n$ \hy{DF} always has the following structure \cite{littlejohn1986}:
\begin{equation}                                                                         \label{wfgauss}
\psi(q,t) =  
\frac{ 
      {\rm e}^{-\frac{1}{2\hbar} (q - \eta_q) \cdot {\bf \Sigma}_{\mathsf S}
                                 (q - \eta_q)
               +\frac{i}{\hbar} (\eta_p \cdot q - \frac{1}{2}\eta_q \cdot \eta_p)} 
               }
{  { (\pi \hbar)^\frac{n}{4} \sqrt{\det{ ( \mathbf{a} + i \mathbf{b} ) }} }  },                   
\end{equation}
where 
\begin{equation}                                                                         \label{complexcov}
{\bf \Sigma}_{\mathsf S} := [\mathsf I_n - i (\bf{ca\!^{\top}} + \bf{db\!^{\top}})]  
                            ({\bf aa\!^\top} + {\bf bb\!^\top})^{-1} 
\end{equation}
is a $n\times n$ matrix constructed with the $n\times n$ blocks of $\mathsf S$ in (\ref{smdef}), $\eta_q := \langle \psi | \hat q | \psi \rangle \in \mathbb{R}^n$ and $\eta_p := \langle \psi | \hat p | \psi \rangle \in \mathbb{R}^n$ are, respectively, the mean value column vectors of position and momentum operators, see below. 

Under the quadratic Hamiltonian, which generates $\mathsf S_t$ in (\ref{sympdyn}), the state in (\ref{wfgauss}) evolves into another Gaussian pure state with the same structure, but with the replacements \cite{littlejohn1986}:
\begin{equation}                                                                         \label{simpev}
{\bf \Sigma}_{\mathsf S} \rightarrow {\bf \Sigma}_{\mathsf{S}_t\mathsf{S}}, \, 
\left(\! \begin{array}{c}
        \eta_q \\
        \eta_p
       \end{array}\!
\right)                  \rightarrow
\mathsf S_t 
\left(\! \begin{array}{c}
        \eta_q \\
        \eta_p
       \end{array} \!     \right)  
+ \int_{0}^{t} \!\!\! dt' {\mathsf S}_{t'} 
\mathsf J \!
\left(\! \begin{array}{c}
        \xi_q \\
        \xi_p
       \end{array}\!    \right). 
\end{equation}
Since the product $\mathsf{S}_t \mathsf{S}$ is a member of the symplectic group, the generic structure in (\ref{wfgauss}) is preserved by this temporal evolution.

The polar structure in (\ref{wfpolar}) is readily obtained for the wave function in (\ref{wfgauss}):
\begin{equation}                                                                         \label{ampph}
\begin{split}
\Omega(q,t) &= 
\frac{ 
      \exp\left[-\frac{1}{4} (q - \eta_q) \cdot {\bf V}^{-1}
                                   (q - \eta_q)\right]  }
     {\left[  (2\pi)^n \det {\bf V} \right]^{1/4}},            \\  
S(q,t) &=  \tfrac{1}{2} (q - \eta_q) \cdot 
          {\rm Im} ({\bf \Sigma}_{\sf S}) (q - \eta_q)  \\
       &+ (\eta_p \cdot q - \tfrac{1}{2}\eta_q \cdot \eta_p)   
 -\tfrac{\hbar}{2} \, {\rm Arg}[ \det( {\bf a} + i {\bf b} ) ].                         
\end{split}
\end{equation}
In these last equations, the matrix $\bf V$ is the position covariance matrix in (\ref{cm}) and, for the Gaussian state in (\ref{wfgauss}), is equal to
\begin{equation}                                                                         \label{complexcov2}
{\bf V} = \frac{\hbar}{2} \left[ {\rm Re} ({\bf \Sigma}_{\sf S}) \right]^{-1} 
        = \frac{\hbar}{2} ({\bf aa\!^\top} + {\bf bb\!^\top}),   
\end{equation}
and is determined by (\ref{complexcov}). Furthermore, the already defined mean values are written as
\[
\begin{split}
\eta_q &= \langle \psi | \hat q | \psi \rangle = 
\int_{\mathbb R^{n}} \!\!\!\!\! \dd^n \! q \,\, [\Omega(q,t)]^2 \, q.  \\ 
\eta_p &= \langle \psi | \hat p | \psi \rangle  = 
\int_{\mathbb R^{n}} \!\!\!\!\! \dd^n \! q \,\, [\Omega(q,t)]^2 \, \partial_q S(q,t).
\end{split}
\]
The mean value of the momenta vector is also in accordance with (\ref{clmom}), {\it i.e.}, 
$p_\text{c} = \eta_p$ for a Gaussian state.  

%%%%%%%%%%%%%%%%%%%%%%%%%%%%%%%%%%%%%%%%%%%%%%%%%%%%%%%%%%%%%%%%%%%%%%%%%%%%%%%%%%%%%%%%%
\section{Convex Decomposition of 
\texorpdfstring{$\widetilde{\mathbf V}_{\! \text{nc}}$}{Vnc}} \label{cdV} %%%%%%%%%%%%%%%
%%%%%%%%%%%%%%%%%%%%%%%%%%%%%%%%%%%%%%%%%%%%%%%%%%%%%%%%%%%%%%%%%%%%%%%%%%%%%%%%%%%%%%%%%
In this Appendix the reader will find the demonstration that the classical/quantum correlations of a mixed states can be written as a convex sum. In summary, we will show the relation among the matrix 
$\widetilde{\mathbf V}_{\! \text{nc}}$ in Eqs.(\ref{correlmix}) and the matrices $\widetilde{\mathbf V}_{\! \text{nc}}^{(k)}$ in Eq.(\ref{correlmix2}). To this end we will only properly rewrite all the derivatives 
appearing in (\ref{correlmix}).

For a question of compactness, let us define $\rho_{qq'} := \langle q | \hat \rho | {q' } \rangle$ for the matrix element in (\ref{denspol}), and by the hermiticity of $\hat \rho$, it is clear that $\rho_{q'\!q} = \rho_{qq'}^\ast$. Using the spectral decomposition (\ref{spectral}), we calculate
\begin{equation}\label{apB1}
\begin{split}
& \rho_{qq} = \bar\Omega(q,q,t) =  \sum_{k}\omega_k |\psi_k^\ast(q)|^2 
= \sum_{k}\omega_k \, \Omega_k^2,  \\
& \partial_j \rho_{qq'}|_{q'=q} = 
\sum_{k}\omega_k \, \psi_k^\ast(q) \, \partial_j \psi_k(q)  \\
&\hspace{1.6cm}= \sum_{k}\omega_k \left[\Omega_k\partial_j\Omega_k + \frac{i}{\hbar} 
                                        \Omega_k^2 \partial_j S_k \right],  \\
\end{split}
\end{equation}
with $\Omega_k = \Omega_k(q,t)$ being the amplitude and $S_k = S_k(q,t)$ the phase of $\psi_k(q) = \langle q|\psi_k\rangle$.

The amplitude in (\ref{denspol}) can be written as 
\[
\bar\Omega(q, {q' }, t ) =  |\rho_{qq'}| = \sqrt{\rho_{qq'}\rho_{q'\!q}}. 
\] 
Taking the derivative with respect to $q_j$ and using (\ref{apB1}), one has
\begin{equation}\label{apB2}
\begin{split}
\partial_j \bar\Omega|_{q'=q} &=  
\left[ \frac{1}{2\bar\Omega} 
       \left( \rho_{q'\!q}\partial_j \rho_{qq'} + 
              \rho_{qq'}\partial_j \rho_{q'\!q}   \right) \right]_{q'=q} \\
&= \tfrac{1}{2} 
\left[\partial_j \rho_{qq'} + \partial_j \rho_{q'\!q}  \right]_{q'=q} 
 \\
&= \tfrac{1}{2} 
\sum_{k}\omega_k \, \partial_j |\psi_k(q)|^2 
= \tfrac{1}{2} \partial_j \bar \Omega(q,q,t). 
\end{split}
\end{equation}
Note the factor $1/2$ at the end, which shows the noncom\-mu\-ta\-ti\-on of the derivative with the selection of the diagonal term $q'=q$. 

A clever way to obtain the derivative of the phase of (\ref{denspol}) is to write it as follows
\begin{equation} \label{apB3}
\begin{split}
\partial_j \bar S|_{q'=q} &=  -i \hbar \,  
\left[  {\rm e}^{- \frac{i}{\hbar} \bar S(q,q',t) } \, 
        \partial_j {\rm e}^{\frac{i}{\hbar} \bar S(q,q',t)} \right]_{q'=q} \\
&= -i \hbar \,{\rm e}^{- \frac{i}{\hbar} \bar S(q,q,t) } \, 
\partial_j \!\! \left.\left(\frac{\rho_{qq'}}{\bar \Omega}\right)\right|_{q'=q}  \\
&= \frac{1}{\bar \Omega(q,q,t)} \sum_k\omega_k \, \Omega_k^2 \, \partial_j S_k, 
\end{split} 
\end{equation} 
where we used Eqs.(\ref{denspol}), Eq.(\ref{apB1}) and that $\bar S(q,q'=q,t) = 0$. Note that 
\[
\left\langle \partial_j \bar S|_{q'=q} \right\rangle_\rho = 
\int_{\mathbb R^{n}} \!\!\!\!\! \dd^n \! q \,\, 
\bar\Omega(q,q,t)\,\partial_j \bar S|_{q'=q} = 
\sum_k\omega_k\, \left\langle \partial_j S_k \right\rangle
\]
is the $j^{\text{th}}$ component of the classical momenta in (\ref{clmomixed}). 

Now, we calculate the second derivative of the amplitude and write it as
\begin{equation}\label{apB5}
\partial^2_{ij} \bar\Omega|_{q'=q}  = \Phi_1(q,t) + \Phi_2(q,t) + \Phi_3(q,t).  
\end{equation}
The first term in the above equation is 
\[
\begin{split}
\Phi_1(q,t) :=& 
- \left[ \frac{\partial_{i}\bar\Omega}{2\bar\Omega^2}   
         \left( \rho_{q'\!q} \, \partial_j \rho_{qq'} + 
                \rho_{qq'}\, \partial_j \rho_{q'\!q}   \right) \right]_{q'=q} \\
=& - \frac{1}{4\bar\Omega(q,q,t)} 
     \partial_i \bar\Omega(q,q,t)\, \partial_j \bar\Omega(q,q,t), 
\end{split}
\] \\
where we used (\ref{apB2}) and the derivatives of $\rho_{qq'}$ were calculated using (\ref{denspol}). The second term is
\[
\begin{split}
\Phi_2(q,t) :=& 
\left[ \frac{1}{2\bar\Omega} 
       \left( \rho_{q'\!q} \, \partial^2_{ij} \rho_{qq'} +  
              \rho_{qq'} \, \partial^2_{ji} \rho_{q'\!q}   \right) \right]_{q'=q} \\                
=& \sum_k\omega_k \, {\rm Re}\!\left[ \psi_k^\ast(q) \, 
                                      \partial^2_{ij} \psi_k(q) \right] \\
=& \sum_{k} \omega_k 
\Omega_k \partial^2_{ij}\Omega_k - 
       \frac{1}{\hbar^2} \sum_{k} \omega_k \Omega_k^2\, \partial_i S_k \, \partial_j S_k , \\
\end{split}
\]
where we use Eq.(\ref{apB2}) and the second derivatives were performed directly from Eq.(\ref{spectral}). The last term is 
\[
\begin{split}
\Phi_3(q,t) :=& 
\left[ \frac{1}{2\bar\Omega} 
       \left( \partial_{i} \rho_{qq'} \, \partial_{j} \rho_{q'\!q} +
              \partial_{i} \rho_{q'\!q} \, \partial_{j} \rho_{qq'}   
              \right) \right]_{q'=q}                                                     \\
=& -\Phi_1(q,t) + \frac{1}{\hbar^2} \bar \Omega(q,q,t)\, 
\partial_{i}\bar S|_{q'=q} \, \partial_{j}\bar S|_{q'=q},
%
% =& -\Phi_1(q,t) + \frac{1}{\hbar^2} \bar P_i \bar P_j,
\end{split} 
\]
where the derivatives were calculated using (\ref{denspol}). Finally, comparing $\widetilde{\mathbf V}_{\! \text{nc}}$ in (\ref{correlmix}) with (\ref{apB5}), we obtain the desired result: 
\[
\!\widetilde{\mathbf V}_{\! \text{nc}} = 
-\hbar^2 \!\sum_{m=1}^3 \!\int_{\mathbb R^{n}} \!\!\! \dd^n \! q \,\, \Phi_m(q,t) 
= \!\sum_{k}\omega_k \widetilde{\mathbf V}_{\! \text{nc}}^{(k)} + \,
                     \delta \widetilde{\bf V}_{\! \text{nc}},  
\]
where $\Phi_1$ cancels with part of $\Phi_3$, while the first summation in the final form of $\Phi_2$ gives rise to the summation of the quantum potentials in (\ref{correlmix2}). The remaining terms of $\Phi_2$ and $\Phi_3$ are grouped in 
\begin{widetext}
\begin{equation}\label{apB7}
\begin{split}
\delta \widetilde{\bf V}_{\!\text nc} :=& 
\sum_{k} \omega_k \! 
\int_{\mathbb R^{n}} \!\!\!\!\! \dd^n \! q \,  
\Omega_k^2 \, \partial_q S_k  (\partial_q S_k)^{\!\top}  
- \int_{\mathbb R^{n}} \!\!\!\!\! \dd^n \! q \, 
\bar \Omega(q,q,t)\, \partial_{q}\bar S|_{q'=q} \, (\partial_{q}\bar S|_{q'=q})^\top \\
%
% =& \sum_{k}\omega_k 
% \int_{\mathbb R^{n}} \!\!\!\!\! \dd^n \! q \, 
% \Omega_k^2 \, \left(\partial_q S_k - \frac{\bar P}{\sqrt{\bar\Omega(q,q,t)}}\right)\! 
%            \left(\partial_q S_k - \frac{\bar P}{\sqrt{\bar\Omega(q,q,t)}}\right)^{\!\!\top}.\\ 
=& \sum_{k}\omega_k \!
\int_{\mathbb R^{n}} \!\!\!\!\! \dd^n \! q \, 
\Omega_k^2 \left(\partial_q S_k - \partial_{q}\bar S|_{q'=q} \right)\!\! 
           \left(\partial_q S_k - \partial_{q}\bar S|_{q'=q} \right)^{\!\!\top}.          
\end{split}
\end{equation}
From its structure, the matrix $\delta {\bf V}_{\!\text{nc}}$ is a positive-semidefinite matrix, which is the most important observation of this appendix. 
\end{widetext}

\

\vspace{10cm}

\

\newpage
%%%%%%%%%%%%%%%%%%%%%%%%%%%%%%%%%%%%%%%%%%%%%%%%%%%%%%%%%%%%%%%%%%%%%%%%%%%%%%%%%%%%%%%%%
\bibliography{Bibliografia}
%%%%%%%%%%%%%%%%%%%%%%%%%%%%%%%%%%%%%%%%%%%%%%%%%%%%%%%%%%%%%%%%%%%%%%%%%%%%%%%%%%%%%%%%%

\end{document}